\documentclass[12pt]{article}
\usepackage{amsmath,amssymb,amsthm,bbm,natbib,color,hyperref}
\usepackage[margin=1.3in]{geometry}
\usepackage{tikz}
\usetikzlibrary{decorations.pathreplacing}
\usepackage{caption}
\usepackage{graphicx}
\usepackage{subcaption}
\usepackage{afterpage}
\usepackage[hang,flushmargin]{footmisc}

\newtheorem{theorem}{Theorem}

\newtheorem{remark}{Remark}
\newcommand{\R}{\mathbb{R}}
\newcommand{\eps}{\epsilon}
\newcommand{\EE}[1]{\mathbb{E}\left[{#1}\right]}
\newcommand{\EEst}[2]{\mathbb{E}\left[{#1}\  \middle| \ {#2}\right]}

\newcommand{\PP}[1]{\mathbb{P}\left\{{#1}\right\}}
\newcommand{\PPst}[2]{\mathbb{P}\left\{{#1}\  \middle| \ {#2}\right\}}

\newcommand{\eqd}{\stackrel{\textnormal{d}}{=}}
\newcommand{\One}[1]{{\mathbbm{1}}\left\{{#1}\right\}}
\newcommand{\qhp}{\widehat{q}^{\,{}_+}}
\newcommand{\qhm}{\widehat{q}^{\,{}_-}}
\newcommand{\alg}{\mathcal{A}}
\newcommand{\Ch}{\widehat{C}}
\newcommand{\muh}{\widehat{\mu}}
\newcommand{\mut}{\widetilde{\mu}}
\newcommand*\samethanks[1][\value{footnote}]{\footnotemark[#1]}
\newcommand{\suppmat}[1]{{#1}}

\title{Predictive inference with the jackknife+}
\author{Rina Foygel Barber\thanks{Department of Statistics, University of Chicago} , 
Emmanuel J.~Cand{\`e}s\thanks{Departments of Statistics and Mathematics, Stanford University} ,
 \\ Aaditya Ramdas\thanks{Department of Statistics and Data Science, Carnegie Mellon University} , 
 Ryan J.~Tibshirani\samethanks \,\! \thanks{Machine Learning Department, Carnegie Mellon University}}
\date{\today}

\begin{document}
\maketitle
\begin{abstract}  This paper introduces the {\em jackknife+}, which is a novel method
  for constructing predictive confidence intervals. Whereas the
  jackknife outputs an interval centered at the predicted response of a
  test point, with the width of the interval determined by the quantiles of leave-one-out
  residuals, the jackknife+ also uses the leave-one-out predictions at the
  test point to account for the variability in the fitted regression function. 
  Assuming exchangeable training samples, we
  prove that this crucial modification permits rigorous coverage
  guarantees regardless of the distribution of the data points, for any algorithm that
  treats the training points symmetrically. Such guarantees are not
  possible for the original jackknife and we demonstrate examples where
  the coverage rate may actually vanish. Our theoretical and empirical
  analysis reveals that the jackknife and the jackknife+ intervals
  achieve nearly exact coverage and have similar lengths whenever the
  fitting algorithm obeys some form of stability. Further, we extend the
  jackknife+ to $K$-fold cross validation and similarly establish
  rigorous coverage properties. Our methods are related to {\em
    cross-conformal prediction} proposed by \citet{vovk2015cross} and
  we discuss connections.\end{abstract}

\section{Introduction}\label{sec:intro}
Suppose that we have i.i.d.~training data $(X_i,Y_i)\in\R^d\times\R$, $i=1,\dots,n$, and a new test point $(X_{n+1},Y_{n+1})$ drawn independently from the same distribution. We would like to fit a regression model to the training data, i.e., a function $\muh:\R^d\rightarrow\R$ where $\muh(x)$ predicts $Y_{n+1}$ given a new feature vector $X_{n+1}=x$, and then provide a prediction interval for the test point---an interval around $\muh(X_{n+1})$ that is likely to contain the true test response value $Y_{n+1}$. 
Specifically, given some target coverage level $1-\alpha$, we would like to construct a prediction interval $\Ch_{n,\alpha}$
as a function of the $n$ training data points, such that
\[\PP{Y_{n+1}\in \Ch_{n,\alpha}(X_{n+1})} \geq 1-\alpha,\]
where the probability is taken with respect to a new test point $(X_{n+1},Y_{n+1})$ as well as with respect to the training data.

A naive solution might be to use the residuals on the training data, $|Y_i - \muh(X_i)|$, to estimate the typical prediction error 
on the new test point---for instance, we might consider the prediction interval
\begin{equation}\label{eqn:naive_intro}\muh(X_{n+1})\pm \textnormal{\Big(the $(1-\alpha)$ quantile of $|Y_1 - \muh(X_1)|,\dots,|Y_n - \muh(X_n)|$\Big)}.\end{equation}
However, in practice, this interval would typically undercover (meaning that the probability that $Y_{n+1}$ lies in this interval 
would be lower than the target level $1-\alpha$), since due to overfitting,
the residuals on the training data points $i=1,\dots,n$ are typically smaller than the residual on the previously unseen test point, i.e., $|Y_{n+1}-\muh(X_{n+1})|$.

In order to avoid the overfitting problem, the jackknife prediction method computes a margin of error with a leave-one-out construction: 
\begin{itemize}
\item For each $i=1,\dots,n$, fit the regression function $\muh_{-i}$ to the training data with the $i$th point removed, and
compute the corresponding leave-one-out residual, $ |Y_i - \muh_{-i}(X_i)|$.
\item Fit the regression function $\muh$ to the full training data, and output the prediction interval
\end{itemize}
\begin{equation}\label{eqn:jackknife_intro}
\muh(X_{n+1})\pm \textnormal{\Big(the $(1-\alpha)$ quantile of $|Y_1 - \muh_{-1}(X_1)|,\dots,|Y_n - \muh_{-n}(X_n)|$\Big)}.\end{equation}
Intuitively, this method 
should have the right coverage properties on average since it avoids overfitting---the leave-one-out residuals $|Y_i - \muh_{-i}(X_i)|$ reflect the typical magnitude of the error in predicting a new data point after fitting to a sample size $n$ (or, almost equivalently, $n-1$), unlike the naive
method where the residuals on the training data are likely to be too small due to overfitting.

However, the jackknife procedure does not have any universal theoretical guarantees. Although many results are known under asymptotic settings or under assumptions of stability of the regression algorithm $\muh$ (we will give an overview below), it is nonetheless the case that, for settings where $\muh$ is unstable, the jackknife method may lose predictive coverage---for example, we will see in our simulations in Section~\ref{sec:empirical} that the jackknife can have extremely poor coverage using least squares regression when the sample size $n$ is close to the dimension $d$.

In this paper, we introduce a new method, the {\em jackknife+}, that provides non-asymptotic coverage guarantees under no assumptions beyond the training and test data being exchangeable. 
We will see that the jackknife+ offers, in the worst case, a $1-2\alpha$ coverage rate (where $1-\alpha$ is the target), while the original jackknife may even have zero coverage
in degenerate examples. On the other hand, empirically we often observe that the two methods yield nearly identical
intervals and both achieve $1-\alpha$ coverage. Theoretically, we will see that under a suitable notion of stability,  
the jackknife+ and jackknife both provably yield close to $1-\alpha$
coverage.

\subsection{Background}
The idea of resampling or subsampling from the available data, in order to assess
the accuracy of our parameter estimates or predictions, has a rich history in the statistics literature.
Early works developing the jackknife and bootstrap methods include \citet{quenouille1949approximate,quenouille1956notes,tukey1958bias,miller1974jackknife,efron1979bootstrap,stine1985bootstrap}.
Several papers from this period
include leave-one-out methods for assessing or calibrating predictive accuracy, similar to the predictive interval constructed
in~\eqref{eqn:jackknife_intro} above, e.g., \citet{stone1974cross,geisser1975predictive,butler1980predictive},
generally using the term ``cross-validation'' to refer to this approach.
(In this work, we will instead use the term ``jackknife'' to refer to the leave-one-out style of prediction methods, as is
common in the modern literature.)
\citet{efron1983leisurely} provides an overview of the early
literature on these types of methods.

While this rich literature demonstrated extensive evidence of the reliable
performance of the jackknife
in practice, relatively little has been known about the theoretical properties of this type of method until recently.
\citet{steinberger2016leave,steinberger2018conditional} have developed results proving
valid predictive coverage of the jackknife under assumptions of algorithmic stability, meaning
that the fitted model $\muh$ and its leave-one-out version $\muh_{-i}$ are required to give similar
predictions at the test point. This work builds on earlier results by \citet{bousquet2002stability},
which study generalization bounds for risk minimization through the framework of stability conditions;
an earlier work in this line is that of \citet{devroye1979distribution}, which give analogous results for classification
risk.

In contrast to cross-validation methods, which perform well but are difficult to analyze theoretically, we can instead consider a simple {\em validation} or {\em holdout}
method. We first partition the training data as $\{1,\dots,n\} = S_{\textnormal{train}}\cup S_{\textnormal{holdout}}$, then
 fit $\muh_{\textnormal{train}}$ on the subset $S_{\textnormal{train}}$ of the training data
and construct
a predictive interval
\begin{equation}\label{eqn:holdout}
\muh_{\textnormal{train}}(X_{n+1})\pm  \textnormal{\Big(the $(1-\alpha)$ quantile of $|Y_i - \muh_{\textnormal{train}}(X_i)|$, $i\in S_{\textnormal{holdout}}$\Big)}.
\end{equation}
\citet{papadopoulos2008inductive,vovk2012conditional,lei2018distribution} study this type of method, under the name {\em split conformal prediction} or {\em inductive
conformal prediction}, through the framework of exchangeability, and prove that $1-\alpha$ predictive coverage holds with  no assumptions
on the algorithm $\alg$ or on the distribution of the data (with a small correction to the definition of the quantile). 
This method is also computationally very cheap, as we only need to fit a single regression function $\muh_{\textnormal{train}}$---in contrast, jackknife
and cross-validation methods require running the regression many times. However, these benefits come at a statistical cost. If the training size $|S_{\textnormal{train}}|$ is much smaller than $n$,
then the fitted model $\muh_{\textnormal{train}}$ may be a poor fit, leading to wide prediction intervals; if instead we decide to take $|S_{\textnormal{train}}|\approx n$
then instead $|S_{\textnormal{holdout}}|$ is very small, leading to high variability. 

Finally, \citet{vovk2015cross,vovk2018cross} proposed the {\em cross-conformal prediction} method, which is closely related to the jackknife+. 
We describe the cross-conformal method, and the previously known theoretical guarantees, in detail later on.
Their work is based on the conformal prediction method (see \citet{vovk2005algorithmic,lei2018distribution} for background),
which provably achieves distribution-free predictive coverage at the target level $1-\alpha$ but at an extremely
high computational cost.

\subsection{Notation}
Before proceeding, we first define some notation. First, for any values $v_i$ indexed by $i=1,\dots,n$, 
define\footnote{
In defining the quantiles $\qhp_{n,\alpha}$ of the 
 residuals, we use $(1-\alpha)(n+1)$ rather than $(1-\alpha) n$ to correct for the finite sample size---we will
see later on why this correction is natural. For the jackknife, it is perhaps more common 
to see $n$ in place of $n+1$, i.e.,~the residual quantile is defined slightly differently, but for large $n$ the difference is negligible.
Formally, if $\alpha < \frac{1}{n+1}$ and so $(1-\alpha)(n+1) > n$, then we set $\qhp_{n,\alpha}\{v_i\} = \infty$.
}
\[\qhp_{n,\alpha}\{v_i\} = \textnormal{ the $\lceil (1-\alpha) (n+1)\rceil$-th smallest value of $v_1,\dots,v_n$},\]
the $1-\alpha$ quantile of the empirical distribution of these values.
Similarly, we will let $\qhm_{n,\alpha}\{v_i\}$ denote the $\alpha$ quantile of the distribution,
\[\qhm_{n,\alpha}\{v_i\} = \textnormal{ the $\lfloor \alpha (n+1)\rfloor$-th smallest value of $v_1,\dots,v_n$ }= -\qhp_{n,\alpha}\{-v_i\}.\]
With this notation, the ``naive'' prediction interval in~\eqref{eqn:naive_intro} can be defined as
\begin{equation}\label{eqn:naive}\Ch^{\textnormal{naive}}_{n,\alpha}(X_{n+1}) = \muh(X_{n+1})\pm \qhp_{n,\alpha}\big\{|Y_i - \muh(X_i)|\}.\end{equation}

Second, we will write $\alg$ to denote the algorithm mapping a training data set of any size, to the fitted regression function. Formally, $\alg$ is a map from $\cup_{m\geq 1} \big(\R^d\times \R\big)^m$ (i.e., the collection of all training sets of any size $m\geq 1$), to the space of functions $\R^d\rightarrow \R$. For example, 
when $\muh$ is the regression function fitted on the training data $(X_1,Y_1),\dots,(X_n,Y_n)$, we can write
\begin{equation}\label{eqn:def_muh_alg}\muh = \alg\Big((X_1,Y_1),\dots,(X_n,Y_n)\Big).\end{equation}
Similarly, to compute the leave-one-out residuals for the jackknife, we let
\begin{equation}\label{eqn:def_muh_minus_i_alg}\muh_{-i} = \alg\Big((X_1,Y_1),\dots,(X_{i-1},Y_{i-1}),(X_{i+1},Y_{i+1}),\dots,(X_n,Y_n)\Big),\end{equation}
and then the jackknife prediction interval~\eqref{eqn:jackknife_intro} can be written as
\begin{equation}\label{eqn:jackknife}\Ch^{\textnormal{jackknife}}_{n,\alpha}(X_{n+1}) = \muh(X_{n+1})\pm \qhp_{n,\alpha}\big\{R^{\textnormal{LOO}}_i\big\},\end{equation}
where $R^{\textnormal{LOO}}_i = |Y_i - \muh_{-i}(X_i)|$ denotes the $i$th leave-one-out residual.

From this point on, we will assume without comment that $\alg$ satisfies a symmetry condition, namely, $\alg$ must be invariant to reordering the data, i.e.,
\begin{equation}\label{eqn:alg_invariant_reordering}
\alg\Big((X_{\pi(1)},Y_{\pi(1)}),\dots,(X_{\pi(m)},Y_{\pi(m)})\Big) = \alg\Big((X_1,Y_1),\dots,(X_m,Y_m)\Big) \end{equation}
for any sample size $m\geq1$, any points $(X_1,Y_1),\dots,(X_m,Y_m)$, and any permutation $\pi$ of the indices $\{1,\dots,m\}$.

\section{The jackknife+}\label{sec:jackknife+}
\begin{figure}[t] 
\begin{tikzpicture}
\input{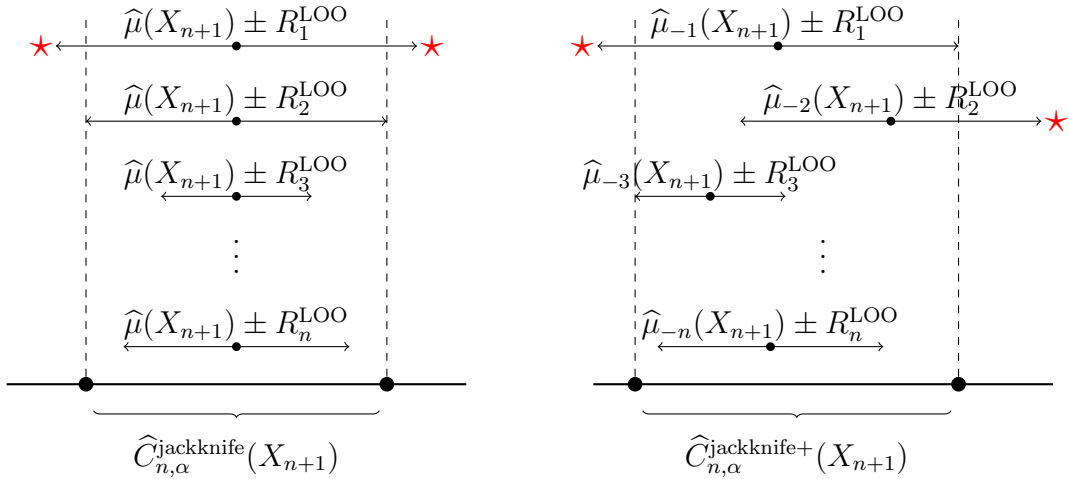}
\end{tikzpicture}
\caption{\small Illustration of the usual jackknife and the new jackknife+. The resulting prediction intervals
 are chosen so that, on either side, the boundary is
exceeded by a sufficiently small proportion of the two sided arrows---above, these are marked with a star.}
\label{fig:jackknife_compare}
\end{figure}

Our jackknife+ method is a modification of the jackknife~\eqref{eqn:jackknife}.
Defining $\muh_{-i}$ as in~\eqref{eqn:def_muh_minus_i_alg}, the jackknife+ prediction interval is given by:
\begin{equation}\label{eqn:jackknife+}
\Ch^{\textnormal{jackknife+}}_{n,\alpha}(X_{n+1}) = {}\\ 
\left[\qhm_{n,\alpha}\big\{\muh_{-i}(X_{n+1}) - R^{\textnormal{LOO}}_i\big\}, \   
\qhp_{n,\alpha}\big\{\muh_{-i}(X_{n+1}) + R^{\textnormal{LOO}}_i\big\}\right].
\end{equation}
To compare this to the usual jackknife, we observe that $\Ch^{\textnormal{jackknife}}_{n,\alpha}(X_{n+1})$ can equivalently be defined as
\[\Ch^{\textnormal{jackknife}}_{n,\alpha}(X_{n+1}) = \left[\qhm_{n,\alpha}\big\{\muh(X_{n+1}) - R^{\textnormal{LOO}}_i\big\}, \  
\qhp_{n,\alpha}\big\{\muh(X_{n+1}) + R^{\textnormal{LOO}}_i\big\}\right].\]
The constructions of the usual jackknife and the new jackknife+ are compared in Figure~\ref{fig:jackknife_compare}.
While both versions of jackknife use the leave-one-out residuals, 
the difference is that for jackknife, we center our interval on the predicted value $\muh(X_{n+1})$ fitted on the full training data,
while for jackknife+ we use the leave-one-out predictions $\muh_{-i}(X_{n+1})$ for the test point.

Figure~\ref{fig:jackknife_compare} illustrates that, if the leave-one-out fitted functions $\muh_{-i}$ are all quite similar
to $\muh$, which was fitted on the full training data, then the two methods should return nearly identical prediction
intervals. On the other hand, in settings where the regression algorithm is extremely sensitive to the training data, such that removing one data point
can substantially change the predicted value at $X_{n+1}$, the output may be quite different. In Section~\ref{sec:stability},
we will examine the role of this type of instability in $\muh$ more closely.

To give one further interpretation of the difference between the two methods,
while the jackknife interval $\Ch^{\textnormal{jackknife}}_{n,\alpha}(X_{n+1})$ is defined as a symmetric interval
around the prediction $\muh(X_{n+1})$ for the test point~\eqref{eqn:jackknife},
the jackknife+ interval can be interpreted as an interval around the {\em median} prediction,
\[\textnormal{Median}\Big(\muh_{-1}(X_{n+1}),\dots,\muh_{-n}(X_{n+1})\Big),\]
which is guaranteed to lie inside $\Ch^{\textnormal{jackknife+}}_{n,\alpha}(X_{n+1})$ for $\alpha\leq \frac{1}{2}$ (in general, however, the jackknife+
interval will not be symmetric around this median prediction).

As detailed in Section~\ref{sec:empirical}, the jackknife and jackknife+ often perform nearly identically in practice (and generally achieve an empirical coverage level
very close to the target $1-\alpha$), but
 in some more challenging examples where the regression algorithm is less stable, the original jackknife may lose coverage while jackknife+ 
still achieves the target coverage level.

Finally, we remark that in
settings where the distribution of $Y| X$ is highly skewed, it may be more natural to consider an asymmetric version
of this method; we consider this extension in Appendix~\ref{sec:extensions}.

\subsection{Assumption-free guarantees}

Remarkably, although the jackknife+ method appears to only be a slight modification of jackknife,
our main result proves that the jackknife+ is guaranteed to achieve predictive 
coverage at the level $1-2\alpha$, without making any assumptions on the distribution of the data $(X,Y)$ or
the nature of the regression method $\alg$.

For this theorem, and all results that follow, all probabilities are
stated with respect to the distribution of the training data points
$(X_1,Y_1),\dots$, $(X_n,Y_n)$ and the test data point $(X_{n+1},Y_{n+1})$ drawn i.i.d.~from an arbitrary distribution $P$, and we assume implicitly that the regression method $\alg$ is invariant to the ordering of the data~\eqref{eqn:alg_invariant_reordering}.
We will treat the sample size $n\geq 2$ and the target coverage level $\alpha\in[0,1]$ as fixed throughout.

\begin{theorem}\label{thm:main}
The jackknife+ prediction interval satisfies
\[\PP{Y_{n+1}\in\Ch^{\textnormal{jackknife+}}_{n,\alpha}(X_{n+1})}\geq 1-2\alpha.\]
\end{theorem}
\noindent This result is proved in Section~\ref{sec:proof_thm:main} using the exchangeability of the $n+1$ data points $(X_1,Y_1),\dots,(X_{n+1},Y_{n+1})$---we 
remark that this theorem actually holds more generally under the assumption that the $n+1$ data points are exchangeable,
with the i.i.d.~assumption as a special case. 

In practice, we generally expect to achieve the target level $1-\alpha$ with either version of the jackknife.
A natural question is whether the factor of 2 appearing in the coverage guarantee for jackknife+ is real, or is merely an artifact of the proof.
We would also want to know whether analogous results may be possible for the original jackknife.

In fact, our next result constructs explicit pathological examples to see that, without making assumptions,
we cannot improve our theoretical guarantee for the jackknife+ (i.e., we cannot remove the factor of 2 appearing in Theorem~\ref{thm:main}),
and no guarantee at all is possible for the jackknife. For completeness, we also construct an example to show zero coverage
for the naive method, although for that method we expect to see undercoverage in practice, not just in pathological examples.
\begin{theorem}\label{thm:worstcase}
For any sample size $n\geq 2$, any $\alpha\in[\frac{1}{n+1},1]$, and any dimension $d\geq 1$, there exists a distribution on $(X,Y)\in\R^d\times \R$ and a regression algorithm $\alg$, such that the predictive coverage of the naive prediction interval~\eqref{eqn:naive} and the jackknife prediction interval~\eqref{eqn:jackknife} satisfy
\[
\PP{Y_{n+1}\in\Ch^{\textnormal{naive}}_{n,\alpha}(X_{n+1})} = \PP{Y_{n+1}\in\Ch^{\textnormal{jackknife}}_{n,\alpha}(X_{n+1})}=0.
\]
Furthermore, if $\alpha \leq \frac{1}{2}$, there exists a distribution on $(X,Y)\in\R^d\times \R$
and a regression algorithm $\alg$, such that the predictive coverage of jackknife+ satisfies 
\[\PP{Y_{n+1}\in\Ch^{\textnormal{jackknife+}}_{n,\alpha}(X_{n+1})}\leq 1 - 2\alpha + 6\sqrt{\frac{\log n}{n}}.\]
\end{theorem}
\noindent The proof of this theorem, and the proofs for all our theoretical results presented below, are deferred to \suppmat{Appendix~\ref{sec:appendix_proofs}}.
The example for the original jackknife is simple---we choose the regression algorithm $\alg$ so that models fitted at sample size $n$
are always less accurate than models fitted at sample size $n-1$. The construction for jackknife+ is substantially more technical, and is similar in spirit
to the example sketched in \citet[Appendix A]{vovk2015cross} for cross-conformal predictors in the setting of exchangeable data.
(The constant 6 on the vanishing term
in the bound for jackknife+ is simply an artifact
of the proof, and can certainly be improved with a more careful construction.)

\subsection{The jackknife-minmax method}

We have seen that the best possible coverage guarantee for jackknife+, in the assumption-free setting, is $1-2\alpha$ rather than the target level $1-\alpha$.
To address this gap, we can consider a more conservative alternative to the jackknife+, 
which will remove the factor of 2 from the theoretical bound. We define the jackknife-minmax method as follows:
\begin{multline}\label{eqn:jackknife-mm}\Ch^{\textnormal{jack-mm}}_{n,\alpha}(X_{n+1}) ={}\\ \left[ \min_{i=1,\dots,n}\muh_{-i}(X_{n+1}) - \qhp_{n,\alpha}\big\{R^{\textnormal{LOO}}_i\big\}, \ \max_{i=1,\dots,n}\muh_{-i}(X_{n+1}) + \qhp_{n,\alpha}\big\{R^{\textnormal{LOO}}_i\big\}\right].\end{multline}
It is simple to verify that this interval is strictly more conservative than jackknife+, meaning that for any data set, we have
\[\Ch^{\textnormal{jackknife+}}_{n,\alpha}(X_{n+1}) \subseteq \Ch^{\textnormal{jack-mm}}_{n,\alpha}(X_{n+1}).\]
The advantage that jackknife-minmax provides is that, without any assumptions on the algorithm or distribution of the data, it always achieves the target coverage rate.
\begin{theorem}\label{thm:minmax}
The jackknife-minmax prediction interval satisfies
\[\PP{Y_{n+1}\in\Ch^{\textnormal{jack-mm}}_{n,\alpha}(X_{n+1})}\geq 1-\alpha.\]
\end{theorem}
\noindent In practice, however, we will see that the jackknife-minmax
prediction interval is generally too conservative.

\section{CV+ for $K$-fold cross-validation}

Suppose that we split the training sample into $K$ disjoint subsets $S_1,\dots,S_K$ each of size $m=n/K$ (assumed to be an integer). 
Let
\[\muh_{-S_k} = \alg\Big((X_i,Y_i): i\in \{1,\dots,n\}\backslash S_k\Big)\]
be the regression function fitted onto the training data with the $k$th subset removed.
To assess the quality of our regression algorithm using cross-validation (CV), we would consider the residuals from this $K$-fold process,
namely, 
\[R^{\textnormal{CV}}_i = \big|Y_i - \muh_{-S_{k(i)}}(X_i)\big|, \ i=1,\dots,n,\]
where $k(i)\in\{1,\dots,K\}$ identifies the subset that contains $i$, i.e., $i\in S_{k(i)}$.
Using these residuals, we can define the CV+ prediction interval as
\begin{equation}\label{eqn:CV+}
\Ch^{\textnormal{CV+}}_{n,K,\alpha}(X_{n+1}) ={}\\ \left[\qhm_{n,\alpha}\big\{\muh_{-S_{k(i)}}(X_{n+1}) - R^{\textnormal{CV}}_i\big\}, \   
\qhp_{n,\alpha}\big\{\muh_{-S_{k(i)}}(X_{n+1}) + R^{\textnormal{CV}}_i\big\}\right].
\end{equation}
Of course, jackknife+ can be viewed as a special case of CV+, by setting $K=n$.
The advantage of the CV+ method, when we choose a smaller $K$, is that we only need to compute $K$ rather than $n$ models---however, this will likely
come at the cost of slightly wider intervals, because the models $\muh_{-S_k}$ are fitted using a lower sample size (i.e., $n(1-1/K)$)
and will lead to slightly larger residuals.

\subsection{Assumption-free guarantee for CV+}
Our next result verifies that the CV+ prediction interval enjoys essentially the same worst-case coverage
guarantee as jackknife+. 
\begin{theorem}\label{thm:CV+}
The $K$-fold CV+ prediction interval satisfies the following coverage guarantees:
\begin{enumerate}
\item[(a)] (Adapted from \citet{vovk2012combining,vovk2018cross}.)
\[\PP{Y_{n+1}\in\Ch^{\textnormal{CV+}}_{n,K,\alpha}(X_{n+1})}\geq 1-2\alpha -\frac{2(1-1/K)}{n/K+1}.\]
\item[(b)]
\[\PP{Y_{n+1}\in\Ch^{\textnormal{CV+}}_{n,K,\alpha}(X_{n+1})}\geq 1-2\alpha - \frac{1-K/n}{K+1}.\]
\end{enumerate}
Combining the two bounds, it follows that for all $K$,
\begin{multline*}\PP{Y_{n+1}\in\Ch^{\textnormal{CV+}}_{n,K,\alpha}(X_{n+1})}\geq 1 - 2\alpha - \min\left\{ \frac{2(1-1/K)}{n/K+1}, \frac{1-K/n}{K+1}\right\} \\ {}\geq  1- 2\alpha - \sqrt{2/n}.\end{multline*}
\end{theorem}
\noindent 
The first part of this result, part (a), is derived from the work \citet{vovk2012combining,vovk2018cross}---we give more
details on this in Section~\ref{sec:crossconf} below. This known result proves that the worst-case coverage is essentially $1-2\alpha$ when $K$ is sufficiently small, i.e., $K\ll n$.
Our new work, proving part (b), completes the picture by giving a meaningful bound for the case where $K$ is large (at the extreme, $K=n$ for leave-one-out methods). By combining the two bounds, we see that coverage is essentially $1-2\alpha$ at {\em any} $K$, since the excess noncoverage
is at most $\sqrt{2/n}$ uniformly over any choice of $K$.

We can also compare our result to the holdout or split conformal method~\eqref{eqn:holdout}, which is equivalent to fitting a model $\muh_{-S_1}$ and constructing
the prediction interval using the quantile of the residuals $R^{\textnormal{CV}}_i$ for $i\in S_1$, but using only a single subset $S_1$ (without repeating $K$ times
for each fold in the partition $S_1,\dots, S_K$). As discussed earlier, this method offers an assumption-free guarantee of $1-\alpha$ coverage,
but this comes at the cost of higher variance due to the single split---in contrast, CV+ reduces variance by averaging over all $K$ splits, but
at the cost of a weaker theoretical guarantee.

\subsection{Related method: cross-conformal predictors}\label{sec:crossconf}
Our proposed CV+ prediction interval
 is related to the {\em cross-conformal prediction} method of \citet{vovk2015cross,vovk2018cross}, which (in its symmetric version)
 returns the predictive set
\begin{multline}\label{eqn:crossconf}
\Ch^{\textnormal{cross-conf}}_{n,K,\alpha}(X_{n+1}) = \Bigg\{y\in \R : {}\\
\frac{\tau + {\displaystyle \sum_{i=1}^n} \One{\big| y - \muh_{-S_{k(i)}}(X_{n+1})\big| < R^{\textnormal{CV}}_i} +\tau\One{\big| y - \muh_{-S_{k(i)}}(X_{n+1})\big| = R^{\textnormal{CV}}_i}  }{n+1} >\alpha\Bigg\}.
\end{multline}
Here $\tau\sim\textnormal{Unif}[0,1]$ introduces randomization into the method.
By comparing to CV+, we can verify that
\begin{equation}\label{eqn:crossconf_vs_CV+}
\Ch^{\textnormal{cross-conf}}_{n,K,\alpha}(X_{n+1}) \subseteq \Ch^{\textnormal{CV+}}_{n,K,\alpha}(X_{n+1})
\end{equation}
deterministically (we demonstrate this in \suppmat{Appendix~\ref{sec:crossconf_app}}). The two methods will sometimes produce the same output, but not always---in particular, $\Ch^{\textnormal{cross-conf}}_{n,K,\alpha}(X_{n+1})$ may
in principle return a predictive set that is a disjoint union of multiple intervals, while CV+ always returns an interval.

We next compare our theoretical findings with the known results for cross-conformal.
\citet{vovk2018cross} show that the $K$-fold cross-conformal method has coverage at least\footnote{\citet{vovk2018cross} do not 
state this coverage result directly, but instead prove $1-2\alpha$ coverage for a modification of the cross-conformal method; however,
these two formulations can be shown to be equivalent. We give details in \suppmat{Appendix~\ref{sec:crossconf_app}}.}
\begin{equation}\label{eqn:crossconf_cov} 1 - 2\alpha -  2(1-\alpha)\frac{1-1/K}{n/K+1}.\end{equation}
When $K$ is small, this additional term is negligible, and so we essentially have $1-2\alpha$ coverage for cross-conformal.
However for large $K$, such as $K=n$ for the leave-one-out method, their earlier result does not yield a meaningful guarantee---the guaranteed coverage level is zero.
 In contrast, our new assumption-free result in Theorem~\ref{thm:main} proves $1-2\alpha$ coverage for the jackknife+ method (i.e., with $K=n$ folds),
and Theorem~\ref{thm:CV+} ensures $1-2\alpha - \sqrt{2/n}$ coverage for $K$-fold CV+ at any choice of $K$. 

\begin{remark}\label{rem:cross-conf}
By examining the proofs of Theorems~\ref{thm:main} and~\ref{thm:CV+}, 
we can see that the arguments apply directly to the $K$-fold (or $n$-fold) cross-conformal method;
that is, our proofs for these theorems also establish that
\begin{multline*}\PP{Y_{n+1}\in\Ch^{\textnormal{CV+}}_{n,K,\alpha}(X_{n+1})}\geq 1 - 2\alpha - \min\left\{ \frac{2(1-1/K)}{n/K+1}, \frac{1-K/n}{K+1}\right\} \\ {}\geq  1- 2\alpha - \sqrt{2/n}\end{multline*}
for $K$-fold cross-conformal with any $K$. In the special case that $K=n$ we are guaranteed coverage  $\geq 1-2\alpha$. 
The first term in the minimum was established by
\citet{vovk2018cross} as presented in~\eqref{eqn:crossconf_cov} above,
but the second term (which allows for meaningful coverage for large values of $K$, e.g., $K=n$) is a new result.\end{remark}

\subsection{An alternative method: conformal prediction}\label{sec:fullconf}
The final related method we present is {\em conformal prediction} \citep{vovk2005algorithmic}.
(We will sometimes refer to this method as ``full'' conformal prediction
in order to distinguish it from the split conformal or cross-conformal methods described above.)
Given the base algorithm $\alg$, the full conformal prediction method outputs a prediction set (which consists of a union of one or more
intervals) constructed as follows:
\begin{equation}\label{eqn:fullconf}
\Ch^{\textnormal{conf}}_{n,\alpha}(X_{n+1}) = \left\{y\in\R : \big|y - \muh^y(X_{n+1})\big| \leq \qhp_{n,\alpha}\big\{\big|Y_i - \muh^y(X_i)\big|\big\} \right\},\end{equation}
where 
\[\muh^y = \alg\Big((X_1,Y_1),\dots,(X_n,Y_n),(X_{n+1},y)\Big)\]
denotes the output of the algorithm run on the training data augmented with the hypothesized test point $(X_{n+1},y)$.
In other words, to determine whether to include a value $y$ in the prediction set at a new point $X_{n+1}$,
we need to train the algorithm on the training+test data (as though $Y_{n+1}=y$ were the true response value), 
and then see whether the residual of the test point ``conforms'' with the
residuals on the remaining $n$ points. 
The exchangeability of the test and training data ensures that
\[\PP{Y_{n+1}\in\Ch^{\textnormal{conf}}_{n,\alpha}(X_{n+1})}\geq  1- \alpha,\]
i.e., coverage at the target level. However, this desirable theoretical property comes at a high cost---we can see by construction
of the prediction interval $\Ch^{\textnormal{conf}}_{n,\alpha}(x)$ that the training algorithm $\alg$ needs to be rerun 
for every test point feature vector $x$ we might consider, and for every possible response value $y\in\R$ (or in practice,
for each $y$ in a fine grid over $\R$).
In certain special cases there are computational tricks allowing for efficient calculation of the prediction set---for example
linear regression or ridge regression~\citep{burnaev2014efficiency}, and the Lasso~\citep{lei2017fast}.
Outside of these special cases, full conformal is prohibitively expensive in practice, even on moderately
sized data sets; while it provides an extremely elegant and theoretically rigorous framework for distribution-free
inference, it is not practical in many applied settings.

\section{Summary of coverage guarantees and computational costs}\label{sec:summary_1}

In light of these theoretical results, which method should a statistician choose in practice?
The following table summarizes the theoretical results behind each of
  the methods under consideration, and the typical empirical performance that
we have observed. 

\begin{table}[h]\centering
\begin{tabular}{l|c|c}
Method&Assumption-free theory&Typical empirical coverage\\
\hline
Naive~\eqref{eqn:naive} & No guarantee &  $< 1-\alpha$\\
 Split conf.~(holdout)~\eqref{eqn:holdout} &  $\geq 1-\alpha$ coverage & $\approx 1-\alpha$\\
Jackknife~\eqref{eqn:jackknife} & No guarantee &  $\approx 1-\alpha$, or  $< 1-\alpha$ if $\muh$ unstable\\
Jackknife+~\eqref{eqn:jackknife+} & $\geq 1-2\alpha$ coverage& $\approx 1-\alpha$\\
Jackknife-minmax~\eqref{eqn:jackknife-mm} & $\geq 1-\alpha$ coverage&  $> 1- \alpha$\\
 Full conformal~\eqref{eqn:fullconf} &  $\geq 1-\alpha$ coverage &  $\approx 1-\alpha$, or $> 1-\alpha$ if $\muh$ overfits\\
 K-fold CV+~\eqref{eqn:CV+} &  $\geq 1-2\alpha$ coverage&  $\gtrsim 1- \alpha$\\
 K-fold cross-conf.~\eqref{eqn:crossconf} &  $\geq 1-2\alpha$ coverage&  $\gtrsim 1-\alpha$\\
\end{tabular}
\end{table}

Given the theoretical and empirical properties of the various options,
we therefore recommend the jackknife+ as a practical alternative to the
usual jackknife predictive intervals.  On the one hand, the empirical
performance of the jackknife+ is nearly identical to that of the
original jackknife (assuming we avoid pathological examples), with
both methods giving intervals of nearly the same width and achieving
close to the target $1-\alpha$ coverage level. However, while the
jackknife offers no theoretical guarantees in the absence of stability
assumptions, the jackknife+ achieves at least $1-2\alpha$ coverage in
the worst possible case. On the other hand, the methods
  achieving $1-\alpha$ (rather than $1-2\alpha$) coverage guarantees
  are either less statistically efficient in the sense of producing
  wider intervals (split conformal uses models fitted on a smaller
  portion of the data while jackknife-minmax is generally too
  conservative), or suffer from computational infeasibility (full
  conformal is computationally prohibitive aside from perhaps a few
  special cases).

  We now turn to a direct comparison of the computational costs of
  these eight methods.  The split conformal and naive methods require
only one run of the regression algorithm $\alg$ (to fit $\muh$ on the
full training data), while each of the jackknife methods requires $n$
runs (to fit $\muh_{-i}$ for each $i=1,\dots,n$---and one additional
run to fit $\muh$, in the case of the original jackknife). If the
training sample size $n$ is so large that fitting $n$ regression
functions is not feasible, we may instead prefer to use $K$-fold
cross-validation. In contrast, the full conformal method must
  train $\alg$ many more times---once for each possible combination of
  a test point feature vector $x$ and a possible response value
  $y$---except for special cases such as linear regression or ridge
  regression.  These observations are summarized below:

\begin{table}[h]\centering
\begin{tabular}{l|c|c}
Method&Model training cost&Model evaluation cost\\
\hline
 Naive~\eqref{eqn:naive} &$1$ &$n +  n_{\textnormal{test}}$ \\
 Split conf.~(holdout)~\eqref{eqn:holdout} &$1$ &\texttt{"}\\
 Jackknife~\eqref{eqn:jackknife} & $n$ &\texttt{"}\\
 Jackknife+~\eqref{eqn:jackknife+} & $n$ &$ n_{\textnormal{test}}\cdot n$\\
 Jackknife-minmax~\eqref{eqn:jackknife-mm} &$n$ & \texttt{"} \\
 K-fold CV+~\eqref{eqn:CV+} & $K$ & $n + n_{\textnormal{test}}\cdot K$ \\
 K-fold cross-conf.~\eqref{eqn:crossconf} & $K$ & \texttt{"} \\
 Full conformal~\eqref{eqn:fullconf} & $n_{\textnormal{test}} \cdot n_{\textnormal{grid}}$ & $n_{\textnormal{test}} \cdot n_{\textnormal{grid}} \cdot n$\\
\end{tabular}
\end{table}
This table compares the computational cost (ignoring constants)
  of each method when run on a training sample of size $n$, for
  producing prediction sets on $n_{\textnormal{test}}$ many test
  points. The middle column (``Model training cost'') counts the
  number of times that the model fitting algorithm $\alg$ is run on a
  training data set\footnote{It is worth
    mentioning that for several common regression algorithms, the $n$
    leave-one-out residuals can be obtained without refitting $n$
    times, but by simply reweighting the in-sample training
    residuals. Examples include linear smoothing methods like ordinary
    least squares, kernel ridge regression, kernel smoothing, thin
    plate splines and smoothing splines. Another interesting example
    is random forests, where the $i$-th leave-one-out fit can be
    obtained by simply ignoring all trees containing the $i$-th
    point.}  of size (up to) $n$. The value $n_{\textnormal{grid}}$ denotes
    the number of grid points of possible $y$ values (a fine grid over $\R$), used
    in the construction of the full conformal prediction method~\eqref{eqn:fullconf}. The last column (``Model evaluation cost'') counts the
  number of times we evaluate a fitted model $\muh$ on a single
  new data point.  In most settings, the model training cost is
  dominant---for example, training a neural network is far more costly
  than evaluating the prediction of a trained network on a new
  example. There are important exceptions, however, such as $K$-nearest
  neighbors, where computing a prediction incurs the cost of
  identifying the $K$ neighbors of the test point.

\section{Guarantees under stability assumptions}\label{sec:stability}
Next, we consider how adding stability assumptions---conditions that ensure that the fitted regression function $\muh$ is not too sensitive to perturbations of the training
data set---can improve the theoretical guarantees of the jackknife and its variants.
(For simplicity, we only consider leave-one-out methods, and do not examine $K$-fold cross-validation here.)

\subsection{In-sample and out-of-sample stability}

Fix any $\eps\geq 0$, $\nu\in[0,1]$, any sample size $n\geq 2$, and
any distribution $P$ on $(X,Y)$. We say that a regression algorithm
$\alg$ satisfies $(\eps,\nu)$ out-of-sample stability with respect to
the distribution $P$ and sample size $n$ if, for all $i\in\{1,\dots,n\}$,
\begin{equation}\label{eqn:out_of_sample_stability}\PP{\left| \muh(X_{n+1}) - \muh_{-i}(X_{n+1})\right| \leq \eps}\geq 1 - \nu,\end{equation}
for $\muh$ and $\muh_{-i}$ defined as before in~\eqref{eqn:def_muh_alg} and~\eqref{eqn:def_muh_minus_i_alg}.
The probability above is taken with respect to the distribution of the
data points $(X_1,Y_1),\dots,(X_n,Y_n)$, $(X_{n+1},Y_{n+1})$ drawn i.i.d.~from $P$.
Similarly, $\alg$ satisfies $(\eps,\nu)$ in-sample stability with
respect to the distribution $P$ and sample size $n$ if, for all $i\in\{1,\dots,n\}$,
\begin{equation}\label{eqn:in_sample_stability}\PP{ \left| \muh(X_i) - \muh_{-i}(X_i)\right| \leq \eps}\geq 1 - \nu.\end{equation}
Naturally, since the data points are exchangeable, if~\eqref{eqn:out_of_sample_stability} or~\eqref{eqn:in_sample_stability} holds for any single $i\in\{1,\dots,n\}$
then it holds for all $i\in\{1,\dots,n\}$. These types of conditions appear elsewhere in the literature---for
 example \citet{bousquet2002stability} define similar conditions,
termed ``hypothesis stability'' and ``pointwise hypothesis stability''.

While the out-of-sample and in-sample stability properties may at first appear similar, they are extremely different in practice.
Out-of-sample stability requires that, for a test point that is {\em independent of the training data}, the predicted value does not change
much if we remove one point from the training data. In contrast, in-sample stability requires that, {\em for a point in
the training data set}, the predicted
value does not change much if we remove this point itself from the training data set. In a scenario where the model fitting
algorithm suffers from strong overfitting, we would expect in-sample stability to be very poor, while out-of-sample stability may
still hold---for example, we will see in Section~\ref{sec:KNN} that this is the case for $K$-nearest-neighbor methods.
On the other hand, strongly convex regularization, such as ridge regression, induces both in- and out-of-sample stability \citep[Example 3]{bousquet2002stability}.
This is not the case, however, for sparse regression methods (e.g., $\ell_1$ regularization),
which are proved by \citet{xu2012sparse} to be incompatible with in-sample stability.

\subsection{Summary of stability results}
Before giving the details of our theoretical results, we summarize our findings on the various methods' predictive coverage guarantees, with and without stability assumptions:

\begin{table}[h]\centering
\begin{tabular}{l|c|c|c}
Method&\begin{tabular}{@{}c@{}}Assumption-free\\theory\end{tabular}& \begin{tabular}{@{}c@{}}Out-of-sample\\ stability\end{tabular} & \begin{tabular}{@{}c@{}}In-sample and\\out-of-sample stability\end{tabular}\\
\hline
Naive~\eqref{eqn:naive} & No guarantee &No guarantee & $\approx 1-\alpha$\\
Jackknife~\eqref{eqn:jackknife} & No guarantee & $\approx 1-\alpha$ & $\approx 1-\alpha$\\
Jackknife+~\eqref{eqn:jackknife+} & $ 1-2\alpha$& $\approx 1-\alpha$& $\approx 1-\alpha$\\
Jackknife-minmax~\eqref{eqn:jackknife-mm} & $ 1-\alpha$& $ 1-\alpha$& $ 1-\alpha$
\end{tabular}
\end{table}

\noindent The assumption free results are the same as those discussed in Section~\ref{sec:summary_1},
while the results under stability assumptions are presented next
in Theorems~\ref{thm:out_of_sample_stability} and~\ref{thm:in_sample_stability}.

\subsection{Out-of-sample stability and the jackknife}
We will next prove that out-of-sample stability is sufficient for the jackknife and jackknife+ methods to achieve the target coverage rate,
with a slight modification. Define 
\[\Ch^{\textnormal{jackknife},\eps}_{n,\alpha}(X_{n+1}) = \muh(X_{n+1})\pm \Big( \qhp_{n,\alpha}\big\{R^{\textnormal{LOO}}_i\big\} + \eps \Big),\]
and similarly, define
\[\Ch^{\textnormal{jackknife+},\eps}_{n,\alpha}(X_{n+1}) = \Big[\qhm_{n,\alpha}\big\{\muh_{-i}(X_{n+1}) - R^{\textnormal{LOO}}_i\big\} - \eps, \   
\qhp_{n,\alpha}\big\{\muh_{-i}(X_{n+1}) + R^{\textnormal{LOO}}_i\big\} + \eps\Big],\]
which we refer to as the {\em $\eps$-inflated} versions of the jackknife and jackknife+ predictive intervals.

\begin{theorem}\label{thm:out_of_sample_stability}
Suppose that the regression algorithm $\alg$ satisfies the $(\eps,\nu)$ out-of-sample stability property~\eqref{eqn:out_of_sample_stability} with respect to the data distribution $P$ and the sample size $n$.
Then the $\eps$-inflated jackknife prediction interval satisfies
\[\PP{Y_{n+1}\in\Ch^{\textnormal{jackknife},\eps}_{n,\alpha}(X_{n+1})}\geq 1-\alpha-2\sqrt{\nu}.\]
Similarly, the $2\eps$-inflated jackknife+ prediction interval satisfies
\[\PP{Y_{n+1}\in\Ch^{\textnormal{jackknife+},2\eps}_{n,\alpha}(X_{n+1})}\geq 1-\alpha-4\sqrt{\nu}.\]
\end{theorem}
\noindent (The different amounts of inflation, $\eps$ for jackknife versus $2\eps$ for jackknife+, 
are simply an artifact of the particular definition of out-of-sample stability that we use, and should not be interpreted
as a meaningful difference between these two methods.)

We remark that if we additionally assume that, in the data
distribution, $Y|X$ has a bounded conditional density (for example,
$Y=\mu(X) + \mathcal{N}(0,\sigma^2)$ for some unknown true mean
function $\mu(\cdot)$), then the result of
Theorem~\ref{thm:out_of_sample_stability} is sufficient to ensure that
the (non-inflated) jackknife and jackknife+ intervals achieve close to
target coverage.  The reason is this: if the conditional density of
$Y|X$ is bounded by some constant $c<\infty$, then very little
probability can be captured by inflating the interval. Specifically,
\[\PP{Y_{n+1} \in \Ch^{\textnormal{jackknife},\eps}_{n,\alpha}(X_{n+1}) \backslash \Ch^{\textnormal{jackknife}}_{n,\alpha}(X_{n+1})} \leq 2\eps  c.\]
Combined with the result of Theorem~\ref{thm:out_of_sample_stability}, this proves that
\[\PP{Y_{n+1}\in\Ch^{\textnormal{jackknife}}_{n,\alpha}(X_{n+1})}\geq 1-\alpha-2\sqrt{\nu}- 2\eps c.\]
Similarly, for jackknife+, we have
\[\PP{Y_{n+1}\in\Ch^{\textnormal{jackknife+}}_{n,\alpha}(X_{n+1})}\geq 1-\alpha-4\sqrt{\nu} - 4\eps c.\]

\subsection{In-sample stability and overfitting}
To contrast the scenarios of in-sample and out-of-sample stability, we will next demonstrate that adding the in-sample stability
assumption would in fact be sufficient for the ``naive'' prediction interval, defined earlier in~\eqref{eqn:naive},
to have coverage at roughly the target level. Its $\eps$-inflated version is defined as
\begin{equation}\label{eqn:naive_eps}\Ch^{\textnormal{naive},\eps}_{n,\alpha}(X_{n+1}) = \muh(X_{n+1})\pm  \Big( \qhp_{n,\alpha}\big\{|Y_i - \muh(X_i)| \big\} + \eps\Big).\end{equation}
Recall from Section~\ref{sec:intro} that
 we would typically expect $\Ch^{\textnormal{naive}}_{n,\alpha}(X_{n+1})$ to undercover severely due to the overfitting problem (thus
inspiring the use of the jackknife to avoid this issue), and similarly $\Ch^{\textnormal{naive},\eps}_{n,\alpha}(X_{n+1})$ will also
undercover whenever $\eps$ is too small to correct for overfitting. This is often the case even when out-of-sample stability is satisfied.
With in-sample stability, however, this is no longer the case---in other 
words, the in-sample stability property is essentially assuming that inflation by $\eps$ is sufficient to correct for overfitting.

\begin{theorem}\label{thm:in_sample_stability}
Suppose that the regression algorithm $\alg$ satisfies both
the $(\eps,\nu)$ in-sample stability property~\eqref{eqn:in_sample_stability}
and the  $(\eps,\nu)$ out-of-sample stability property~\eqref{eqn:out_of_sample_stability}
with respect to the data distribution $P$ and the sample size $n$.
Then the naive prediction interval satisfies
\[\PP{Y_{n+1}\in \Ch^{\textnormal{naive},2\eps}_{n,\alpha}(X_{n+1})}\geq 1-\alpha-4\sqrt{\nu}.\]
\end{theorem}

\subsection{Example: $K$-nearest-neighbors}\label{sec:KNN}
To give an illustrative example, consider a $K$-nearest-neighbor ($K$-NN) method.
This style of example is also
considered in \citet[Example 4.1]{steinberger2018conditional}, and was studied earlier
by \citet{devroye1979distribution} in the context of estimating the error
of a classifier, and by \citet{bousquet2002stability} in the context of error in regression.
Given a training 
data set $(X_1,Y_1),\dots,(X_n,Y_n)$ and a new test point $x$, our prediction is 
\[\muh(x) = \frac{1}{K}\sum_{i \in N(x)} Y_i,\]
where $N(x)\subset\{1,\dots,n\}$ is the set of the $K$ nearest neighbors to the test point $x$, i.e.,~the $K$
indices $i$ giving the smallest values of $\|X_i - x\|_2$ (of course, we can replace
the $\ell_2$ norm with any other  metric).
We will assume for simplicity that there are no ties between these distances
(for instance, the $X_i$ points might
be continuously distributed on $\R^d$, or we apply a random tie-breaking rule). 
Now consider out-of-sample stability. Let $N(X_{n+1})$ and $N_{-i}(X_{n+1})$
be the $K$-nearest-neighbor sets for the test point $X_{n+1}$ given the full training data, or the training data with data point $i$
removed, respectively. Then we can easily verify that
\[i \not\in N(X_{n+1}) \ \Leftrightarrow \ N(X_{n+1}) = N_{-i}(X_{n+1}) \ \Rightarrow \ \muh(X_{n+1}) = \muh_{-i}(X_{n+1}).\]
Therefore,
\[\PP{\left| \muh(X_{n+1}) - \muh_{-i}(X_{n+1})\right|=0} \geq \PP{i\not\in N(X_{n+1})}  = 1 - \frac{K}{n},\]
where the last step holds by exchangeability of the $n$ training points. This proves that the $K$-NN
method satisfies $(\eps,\nu)$-out-of-sample stability with $\eps=0$ and $\nu = K/n$.
(In contrast, we cannot hope for a similar argument to guarantee
in-sample stability, since we will always have $i\in N(X_i)$; that is, $X_i$ is one
of its own nearest neighbors---and so in general we will have $\muh(X_i) \neq \muh_{-i}(X_i)$.)

Applying the conclusion of Theorem~\ref{thm:out_of_sample_stability} to this setting, then,
we see that $K$-NN leads to a coverage rate at least
\[\PP{Y_{n+1}\in\Ch^{\textnormal{jackknife}}_{n,\alpha}(X_{n+1})}\geq 1-\alpha-2\sqrt{K/n}\]
for the jackknife, and
\[\PP{Y_{n+1}\in\Ch^{\textnormal{jackknife+}}_{n,\alpha}(X_{n+1})}\geq 1-\alpha-4\sqrt{K/n}\]
for the jackknife+.
These results hold with no assumptions whatsoever on the distribution of the data---and in particular,
we do not need to assume that the $K$-NN prediction is accurate or consistent on the given data.

\subsection{Comparison to existing results}\label{sec:literature_stability}
As mentioned above,
\citet{bousquet2002stability} study stability in the context of generalization bounds 
for regression, with the aim of bounding risk rather than predictive inference.
The predictive accuracy of the jackknife under assumptions of algorithm stability was explored
by \citet{steinberger2016leave} for the linear regression setting,
and in a more general setting by \citet{steinberger2018conditional}.
Their stability assumption (see, e.g., \citet[Definition 1]{steinberger2018conditional})
is essentially equivalent to our out-of-sample stability condition~\eqref{eqn:out_of_sample_stability}.
However, the theory obtained in their work is asymptotic, and relies also on distributional 
assumptions (see \citet[(C1)]{steinberger2018conditional}),
namely, that $Y_i = \EEst{Y_i}{X_i} + \nu_i$ where the noise $\nu_i$ is continuously
distributed and  is independent of $X_i$
(for example, this does not allow for heteroskedasticity). In contrast, our guarantee,
 in Theorem~\ref{thm:out_of_sample_stability}, offers a simple finite-sample coverage guarantee
with no distributional assumptions, requiring only algorithm stability.

\section{Proof of Theorem~\ref{thm:main}}\label{sec:proof_thm:main}
Suppose for a moment that we have access to the test point $(X_{n+1},Y_{n+1})$ as well as the training data. For any indices $i,j\in\{1,\dots,n+1\}$ with $i\neq j$, let $\mut_{-(i,j)}$ define the regression function fitted on the training plus test data, with points $i$ and $j$ removed.
 (We use $\mut$ rather than $\muh$ to remind ourselves that the model is fitted on a subset of the combined training and test data $i=1,\dots,n+1$, rather than a subset of only the training data.) Note that $\mut_{-(i,j)} = \mut_{-(j,i)}$ for any $i\neq j$, and that $\mut_{-(i,n+1)} = \muh_{-i}$ for any $i=1,\dots,n$.

Next, we define a matrix of residuals, $R\in\R^{(n+1)\times(n+1)}$, with entries
\[R_{ij} = \begin{cases} + \infty, & i=j,\\
\big|Y_i - \mut_{-(i,j)}(X_i)\big|,& i\neq j,\end{cases}\]
i.e., the off-diagonal entries represent the residual for the $i$th point when both $i$ and $j$ are left out of the regression.
We also define a comparison matrix, $A\in\{0,1\}^{(n+1)\times(n+1)}$,
with entries
\[A_{ij} = \One{R_{ij} > R_{ji}}.\]
In other words, $A_{ij}$ is the indicator for the event that, when excluding data points $i$ and $j$ from the regression,
data point $i$ has higher residual than data point $j$. Naturally we see that $A_{ij}=1$ implies $A_{ji}=0$, for any $i,j$.
We note that this comparison matrix construction is also examined by \citet[Appendix A]{vovk2015cross}, where it is used
to establish that leave-one-out conformal methods fail to achieve $1-\alpha$ coverage.

Next, we are interested in finding data points $i$ with unusually large residuals---the ones that are
hardest to predict. We will define a set $\mathcal{S}(A)\subseteq\{1,\dots,n+1\}$ of ``strange'' points,\footnote{The authors
thank an anonymous reviewer for suggesting this presentation of the proof.}
\[\mathcal{S}(A)= \left\{i\in\{1,\dots,n+1\} : A_{i \bullet} \geq (1-\alpha)
    (n+1)\right\},\]
where $A_{i \bullet} = \sum_{j=1}^{n+1} A_{ij}$ is the $i$th row sum of $A$. 
In other words, the $i$th point is ``strange'' (i.e., $i\in\mathcal{S}(A)$) if it holds that,
when we compare 
the residual $R_{ij}$ of the $i$th point against residual $R_{ji}$ for the $j$th point (for each $j\neq i$),
the residual $R_{ij}$ for the $i$th point is the larger one, for a sufficiently high fraction of these comparisons.

From this point on, the proof will proceed as follows:
\begin{itemize}
\item Step 1: we will establish deterministically that $|\mathcal{S}(A)|\leq 2\alpha(n+1)$, that is, for any comparison matrix $A$
it is impossible to have more than $2\alpha(n+1)$ many strange points.
\item Step 2: using the fact that the data points are i.i.d.~(or more generally exchangeable), we will show that the probability that
the test point $n+1$ is strange (i.e., $n+1\in\mathcal{S}(A)$) is therefore bounded by $2\alpha$.
\item Step 3: finally, we will verify that the jackknife+ interval can only fail to cover
the test response value $Y_{n+1}$ if $n+1$ is a strange point.
\end{itemize}

\paragraph{Step 1: bounding the number of strange points}
This bound is essentially a consequence of Landau's theorem for tournaments
\citep{landau1953dominance}. For data points $i$ and $j$, we say that data point 
$i$ ``wins'' its game against data point $j$, if $A_{ij}=1$; that is, point $i$
has a higher residual than point $j$, under the corresponding regression 
$\mut_{-(i,j)}$. Note that each strange point $i \in \mathcal{S}(A)$ can lose
against at most $\alpha(n+1)-1$ other strange points---this is because point $i$ must win against
at least  $(1-\alpha)(n+1)$ points in total since it is strange, and  
as we have defined it, point $i$ cannot win against itself.  

Let $s=|\mathcal{S}(A)|$ denote the number of strange points. The key
realization is now that, if we think about grouping each {\em pair} of strange
points by the losing point, then we see that there are at most 
$$ s \cdot \big(\alpha(n+1)-1\big) $$ 
pairs of strange points.  This is because there are at most $s$ unique
possibilities for the loser, and for each such loser, it can only lose against
at most $\alpha(n+1)-1$ other strange points, as argued above.  In other
words, we have established  
$$ \frac{s(s-1)}{2} \leq s \cdot \big(\alpha(n+1)-1\big), $$
and rearranging gives $s \leq 2\alpha(n+1)-1 < 2\alpha(n+1)$, as desired.


\paragraph{Step 2: exchangeability of the data points}
We next leverage the exchangeability of the data points to 
show that, since there are at most $2\alpha(n+1)$ strange points among a total of $n+1$ points,
it follows that the test point has at most $2\alpha$ probability of being strange---this
reasoning uses the exchangeability of the data in exactly the same way
as the conformal prediction literature \citep{vovk2005algorithmic}.

To establish this formally, since the data points $(X_1,Y_1),\dots,(X_{n+1},Y_{n+1})$ are exchangeable and the regression fitting algorithm $\alg$ is invariant to the ordering of the data points (the symmetry condition~\eqref{eqn:alg_invariant_reordering}), 
it follows that $A \eqd \Pi A \Pi^\top$ for any $(n+1)\times(n+1)$ permutation matrix $\Pi$, where $\eqd$ denotes equality in distribution. In particular, for any index $j\in\{1,\dots,n+1\}$, suppose we take $\Pi$ to be any permutation matrix with $\Pi_{j,n+1}=1$ (i.e., corresponding to a permutation mapping $n+1$ to $j$). Then, deterministically, we have
\[ n+1 \in\mathcal{S}(A) \ \Leftrightarrow \ j \in \mathcal{S}\big(\Pi A \Pi^\top\big),\]
and therefore,
\[\PP{n+1 \in\mathcal{S}(A)} = \PP{j\in \mathcal{S}\big(\Pi A \Pi^\top\big)} = \PP{j\in \mathcal{S}(A)}\]
for all $j=1,\dots,n+1$. In other words, if we compare an arbitrary training point $j$
versus the test point $n+1$, these two points are equally likely to be strange, by exchangeability of the data.
We can then calculate
\[
\PP{n+1 \in\mathcal{S}(A)} = \frac{1}{n+1}\sum_{j=1}^{n+1} \PP{j\in\mathcal{S}(A)}= \frac{\EE{|\mathcal{S}(A)|}}{n+1}\leq 2\alpha,\]
where the last step applies the result of Step 1.

\paragraph{Step 3: connecting to jackknife+}
Finally, we need to relate the question of coverage of the jackknife+ interval, to our notion of strange points.
Suppose that $Y_{n+1}\not\in\Ch^{\textnormal{jackknife+}}_{n,\alpha}(X_{n+1})$. 
This means that either
\[Y_{n+1} > \qhp_{n,\alpha}\big\{\muh_{-i}(X_{n+1}) + R^{\textnormal{LOO}}_i\big\},\]
which implies that $Y_{n+1} > \muh_{-j}(X_{n+1}) + R^{\textnormal{LOO}}_j$ for at least $(1-\alpha)(n+1)$ many indices $j\in\{1,\dots,n\}$,
or otherwise
\[Y_{n+1} < \qhm_{n,\alpha}\big\{\muh_{-i}(X_{n+1}) - R^{\textnormal{LOO}}_i\big\},\]
which implies that $Y_{n+1} < \muh_{-j}(X_{n+1}) - R^{\textnormal{LOO}}_j$ for at least $(1-\alpha)(n+1)$ many indices $j\in\{1,\dots,n\}$.
In either case, then, we have
\begin{align*}
(1-\alpha)(n+1)
 &\leq \sum_{j=1}^n \One{Y_{n+1} \not \in \muh_{-j}(X_{n+1}) \pm R^{\textnormal{LOO}}_j}\\
&= \sum_{j=1}^n \One{\big|Y_j - \muh_{-j}(X_j)\big| < \big|Y_{n+1} - \muh_{-j}(X_{n+1})\big|} \\
&= \sum_{j=1}^{n+1} \One{R_{j,n+1} < R_{n+1,j}} 
=\sum_{j=1}^{n+1} A_{n+1,j},
\end{align*}
and therefore $n+1\in\mathcal{S}(A)$, that is, point $n+1$ is strange.
Combining this with the result of Step 2, we have
\[\PP{Y_{n+1}\not\in\Ch^{\textnormal{jackknife+}}_{n,\alpha}(X_{n+1})}\leq \PP{n+1\in\mathcal{S}(A)} \leq 2\alpha.\]

\section{Empirical results}\label{sec:empirical}
In this section, we compare seven methods---naive \eqref{eqn:naive}, jackknife \eqref{eqn:jackknife}, jackknife+ \eqref{eqn:jackknife+}, jackknife-minmax \eqref{eqn:jackknife-mm}, CV+ \eqref{eqn:CV+}, split conformal \eqref{eqn:holdout}, and full conformal \eqref{eqn:fullconf}---on simulated and real data.
Code for reproducing all results and figures is available online.\footnote{\url{http://www.stat.uchicago.edu/~rina/jackknife.html}}

\subsection{Simulations}
We first examine the performance of the various prediction intervals on a simulated example,
using least squares as our regression method. We will see that when the training sample size $n$ is equal or approximately
equal to the dimension $d$, the instability of the least squares method (due to poor conditioning of the $n\times d$ design matrix)
leads to a wide disparity in performance between the various methods. 
This simulation is thus designed to demonstrate the role of stability in the performance of these various methods.

\subsubsection{Data and methods} Our target coverage level is $1-\alpha = 0.9$.
We use training sample size $n=100$,
and repeat the experiment at each dimension $d=5,10,\dots,200$, with i.i.d.~data points $(X_i,Y_i)$ generated as
\[X_i \sim \mathcal{N}(0,I_d)\textnormal{ and }Y_i \mid X_i \sim \mathcal{N}(X_i^\top\beta, 1).\]
The true coefficient vector $\beta$ is drawn as $\beta=\sqrt{10}\cdot u$ for a uniform random unit vector $u\in\R^d$.
The regression method $\alg$ is
simply least squares, with the convention that if the linear system is underdetermined then we take the solution with the lowest $\ell_2$ norm (the limit of ridge regression as the regularization tends to zero).
Specifically, for training data $(X_1,Y_1),\dots,(X_n,Y_n)$, we return
the regression function $\muh(x) =x^\top \widehat\beta$, 
\[
  \widehat{\beta} = X_{\textnormal{mat}}^\dagger Y_{\textnormal{vec}},
\]
where $X_{\textnormal{mat}}$ denotes the $n\times d$ matrix of covariates, $Y_{\textnormal{vec}}$ the vector of
responses, and 
$\dagger$ the Moore--Penrose pseudoinverse.

We then generate 100 test data points from the same distribution, and calculate
the empirical probability of coverage (i.e., the proportion of test points for which the prediction interval computed
at the $X$ value contains the $Y$ value) and the average width of the prediction interval.

\begin{figure}[hp]
\centering
\begin{subfigure}[b]{0.45\textwidth}\centering
\caption{ Average coverage}
\includegraphics[width=\textwidth]{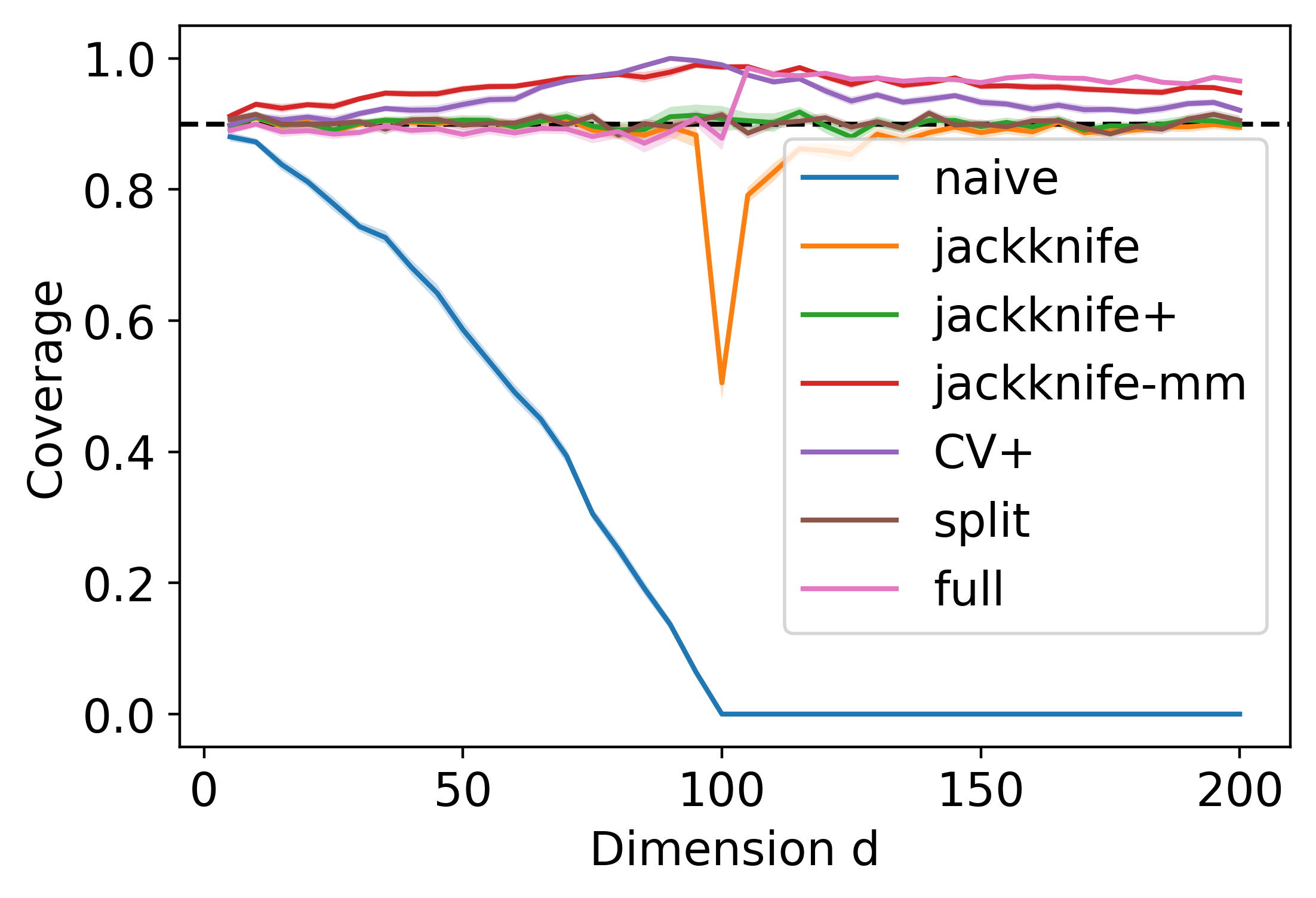}
\end{subfigure}\quad
\begin{subfigure}[b]{0.45\textwidth}\centering
\caption*{(zoomed in)}
\includegraphics[width=\textwidth]{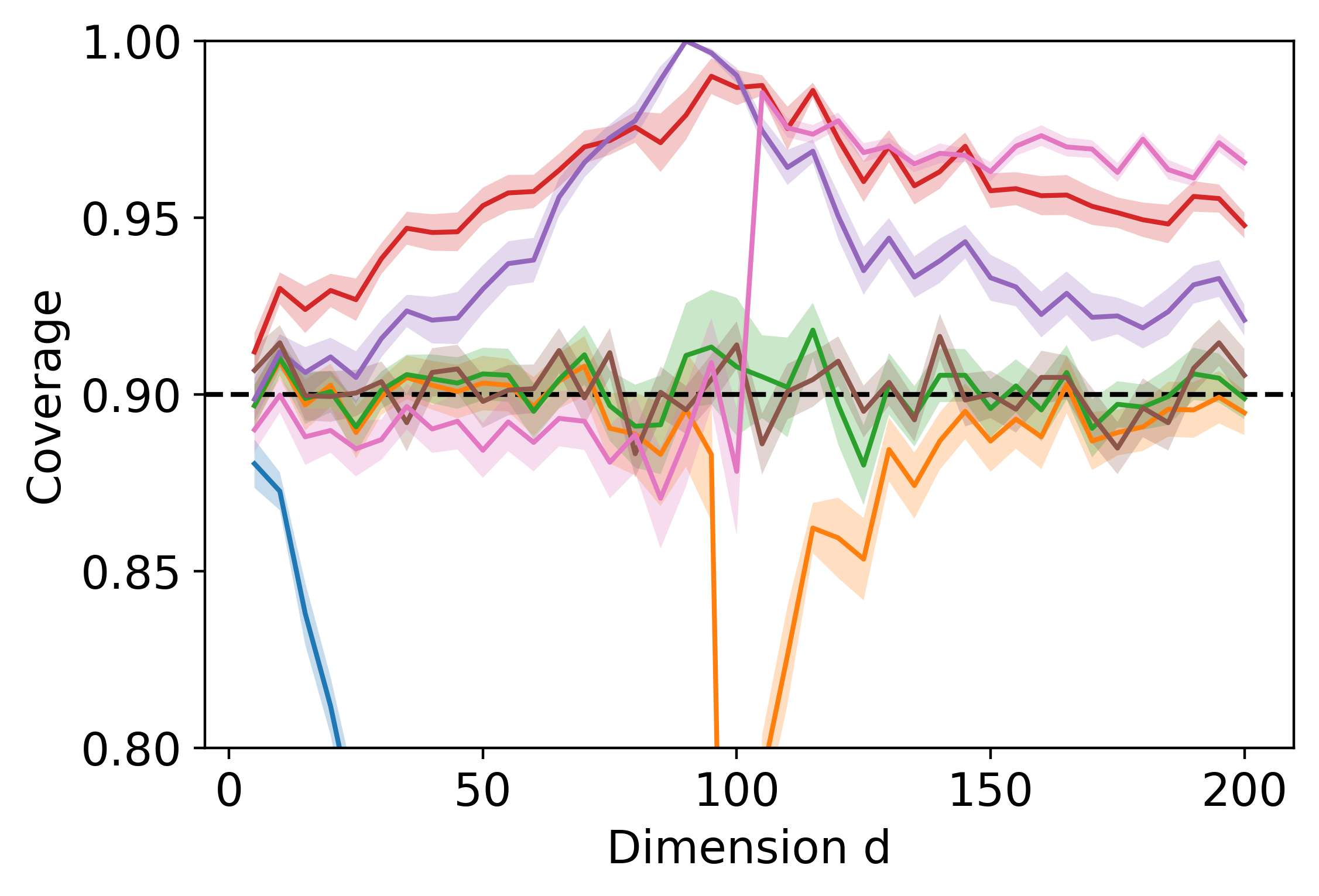}
\end{subfigure}\bigskip\bigskip

\begin{subfigure}[b]{0.45\textwidth}\centering
\caption{Average prediction interval width}
\includegraphics[width=\textwidth]{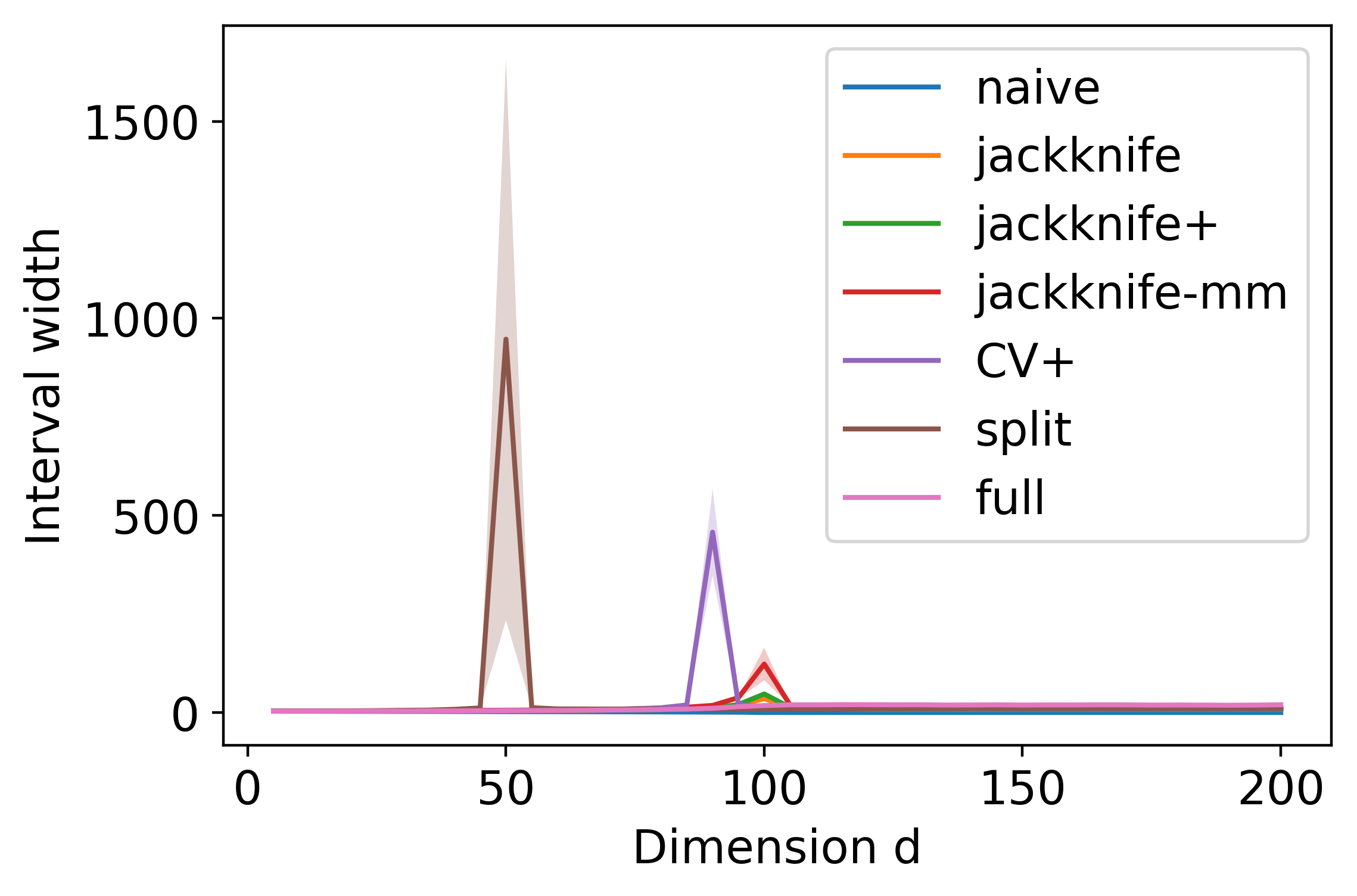}
\end{subfigure}\quad
\begin{subfigure}[b]{0.45\textwidth}\centering
\caption*{(zoomed in)}
\includegraphics[width=0.94\textwidth]{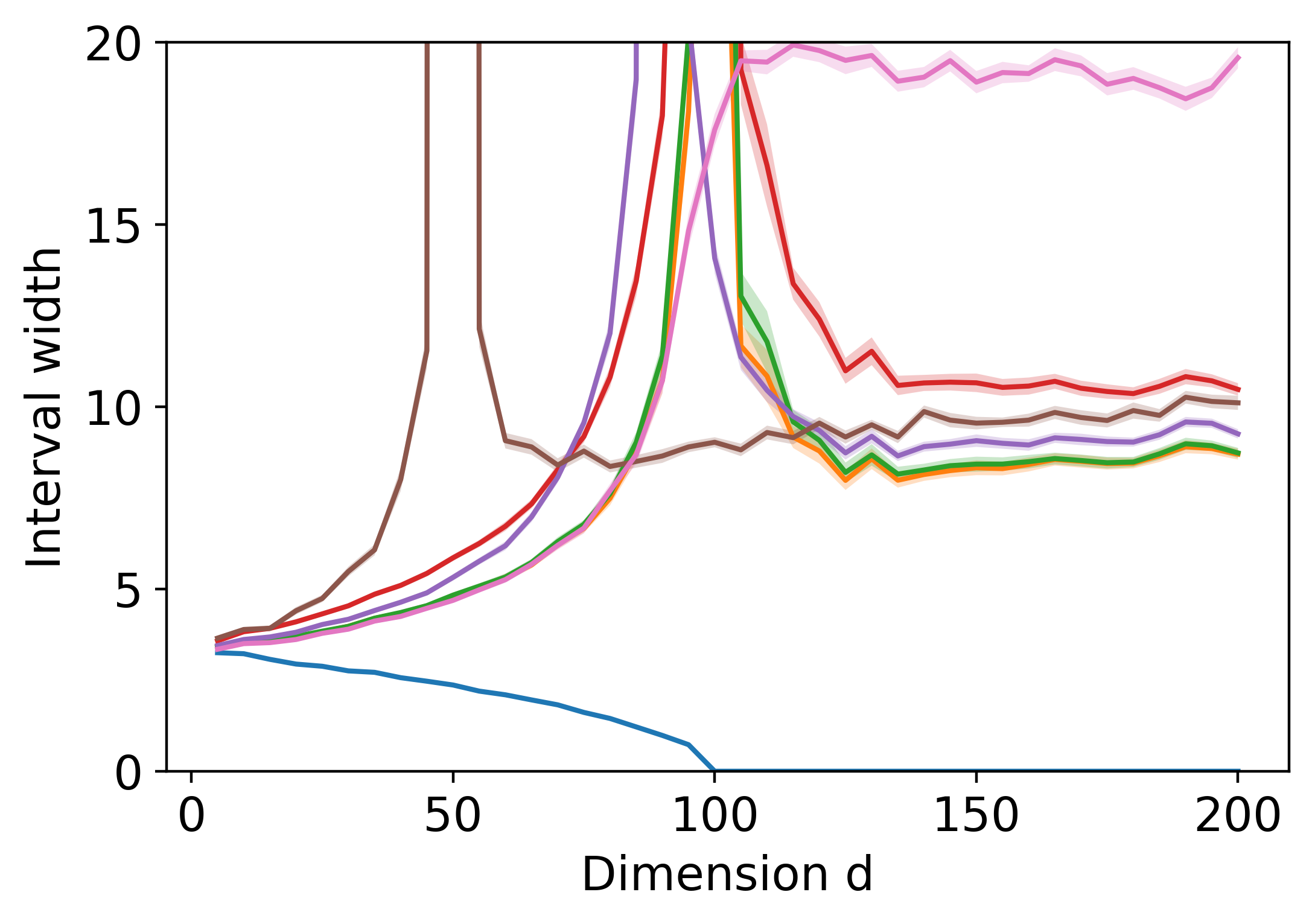}
\end{subfigure}\quad
\caption{Simulation results, showing the coverage and width of the predictive
interval for all methods. The solid lines show the mean over 50 independent trials, with shading to show
$\pm$ one standard error. We observe that the jackknife undercovers around $d=100$ due to instability (since $n=100$).
Jackknife+ and split conformal are the only two methods that maintain the correct
coverage level throughout without under- or over-covering,
but we can observe that jackknife+ often produces shorter intervals
than split conformal. (See text for more details.)}
\label{fig:sim_results}
\end{figure}

\subsubsection{Results}\label{sec:sim_results} Figure~\ref{fig:sim_results} displays the results of the simulation,
averaged over 50 trials (where each trial has an independent draw of the training sample of $n=100$ and the test sample of size 100). 

When $d<n$, the jackknife and jackknife+ show very similar performance, with approximately the right coverage level $1-\alpha = 0.9$ and with
nearly identical interval width. For $d\approx n$ (the regime where least squares is quite unstable), the jackknife has substantial undercoverage---at $d=n$
the jackknife shows coverage rate around $0.5$, and continues to show substantial undercoverage when $d$ is slightly larger than $n$. In this regime, 
the jackknife+ continues to show the right coverage level, at the cost of a prediction interval that is only slightly wider than the
 jackknife. For large $d$, the jackknife
and jackknife+ again show very similar performance. 
In fact, this connects to recent work on interpolation methods (methods that
achieve zero training error). Specifically, \citet{hastie2019surprises} study ``ridgeless'' regression (i.e., the least squares
solution with the lowest $\ell_2$ norm, as in our simulation), and demonstrate that this provides a stable solution with good test error
as long as $d$ is {\em either} sufficiently small or sufficiently large relative to $n$. We see a similar phenomenon in the predictive coverage
performance of the jackknife. 

As expected, the jackknife-minmax is over-conservative, with typical coverage higher than $1-\alpha =0.9$ across all dimensions $d$,
while the naive method drastically undercovers due to increasing overfitting as $d$ grows (and in fact, at $d\geq n$, the training error is exactly zero, 
so the prediction intervals have width zero and coverage zero.)

 When $d>n$, we note that full conformal prediction will always have infinite length intervals since for every potential $y$ in the $(X_{n+1},y)$ pair, all $n+1$ residuals will equal zero. Naturally, in such a situation, full conformal will have coverage equal to one deterministically.  
In practice, it is common to modify the conformal prediction method by truncating
to a finite range, e.g., to the observed range of $Y$ values in the training data (which has minimal effect on the coverage
guarantee \citep{chen2018discretized}); this is why we see finite length intervals for full conformal in our simulation results.

Split conformal is the only method other than jackknife+ to maintain coverage at $0.9$ throughout.
(Note that since split conformal trains on half of the data,
its length spikes near $d=50$, rather than $d=100$ as for the other methods; this is simply due to the 
change in sample size $n/2=50$ used in training. This is a result of instability of OLS when $n \approx d$ and is not reflective of comparisons between holdout and jackknife+.)

\subsection{Real data}
We next compare the various methods on
three real data sets. We will try three regression algorithms: ridge regression, random forests, and neural networks (details given
below). Our aim in these experiments is to demonstrate the typical performance of the various
prediction interval methods in a real data setting; we do not seek to optimize 
the base methods used as our regression algorithms, but are only interested in how
the various prediction interval methods behave in comparison to each other.
Due to the high computational cost of the full conformal method, we do not include it in the comparison.

\subsubsection{Data}
The Communities and Crime data set\footnote{\url{http://archive.ics.uci.edu/ml/datasets/communities+and+crime}} 
\citep{redmond2002data} contains information on 1994 communities, with covariates
such as median income, distribution of ages, family size, etc., and the goal of predicting a response
variable defined as the per capita violent crime rate. After removing categorical variables and variables
with missing data, $d=99$ covariates remain.

The BlogFeedback data set\footnote{\url{https://archive.ics.uci.edu/ml/datasets/BlogFeedback}} 
\citep{buza2014feedback} contains 52397 data points, each corresponding to a single blog post.
The goal is to predict the response variable of the number of comments left on the blog post in the following 24 hours,
using $d=280$ covariates such as the length of the post, the number of comments on previous posts, etc.
Since the distribution of the response is extremely skewed,
we transform it as $Y = \log(\textnormal{1 + \# comments})$. 

The Medical Expenditure Panel Survey 2016 
data set,\footnote{\url{https://meps.ahrq.gov/mepsweb/data_stats/download_data_files_detail.jsp?cboPufNumber=HC-192}}
 provided by 
the Agency for Healthcare Research and Quality,
contains data on individuals' utilization of medical services such as visits to the doctor, hospital stays, etc.
Details on the data collection for older versions of this data set are described in \citet{ezzati2009methodology}.
We select a subset of relevant features, such as age, race/ethnicity, family income, occupation type, etc. 
After splitting categorical features into dummy variables to encode each category separately, 
the resulting dimension is $d=107$.
The goal is to predict the health care system utilization of each individual, which is a composite
score reflecting the number of visits to a doctor's office, hospital visits, days in nursing home care, etc. 
With missing data removed, this data set contains 33005 data points.
Since the distribution of the response is highly skewed,
we transform it as $Y = \log(\textnormal{1 + (utilization score)})$. 

\begin{figure}[hp]
\centering
\begin{subfigure}[h!]{\textwidth}
\caption{Communities and crime data set}
\includegraphics[height=0.18\textheight]{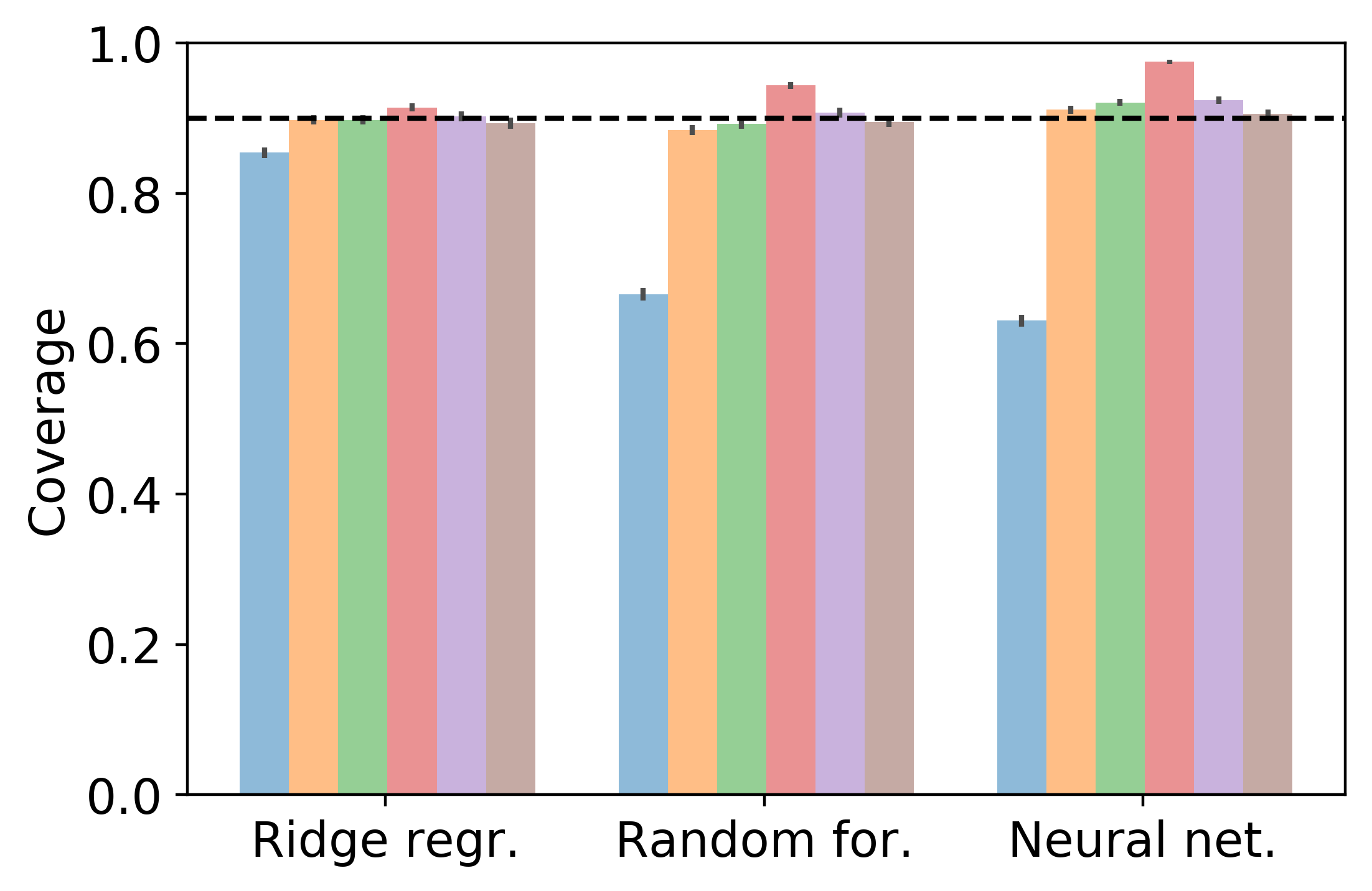}\quad 
\includegraphics[height=0.18\textheight]{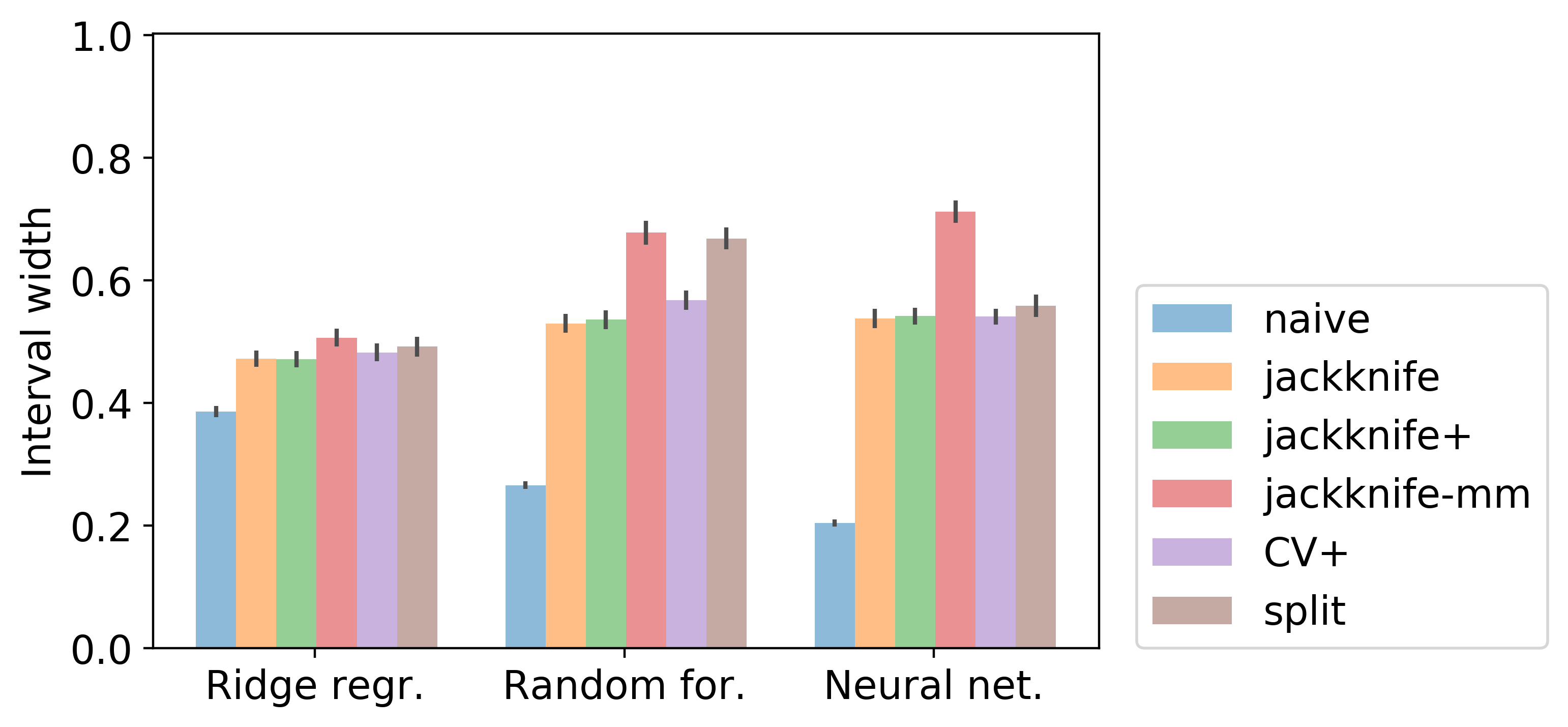}
\label{fig:realdata_results_communities}
\end{subfigure}\smallskip

\begin{subfigure}[h!]{\textwidth}
\caption{BlogFeedback data set}
\includegraphics[height=0.18\textheight]{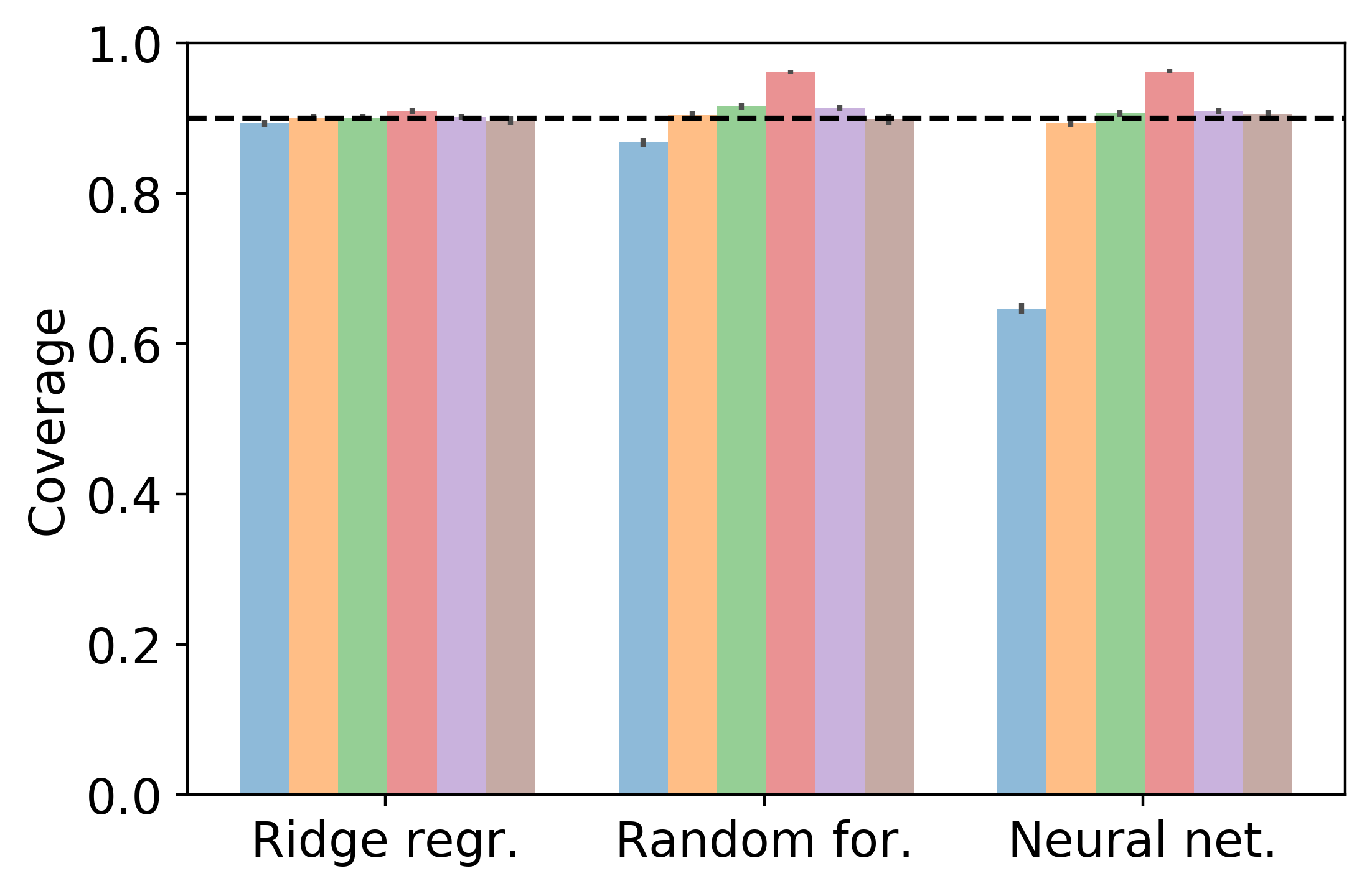}\quad 
\includegraphics[height=0.18\textheight]{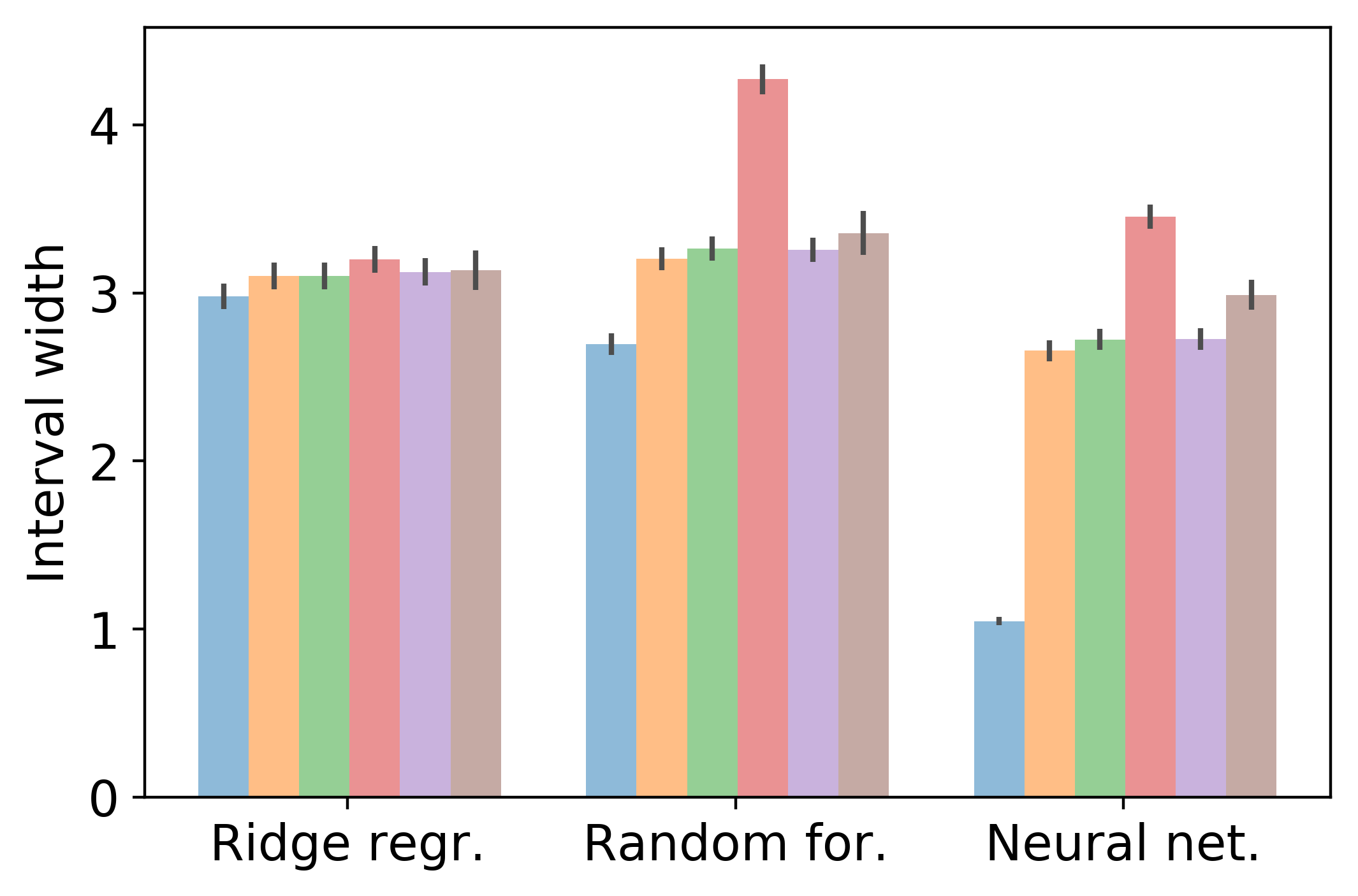}
\label{fig:realdata_results_blog}
\end{subfigure}\bigskip

\begin{subfigure}[h!]{\textwidth}
\caption{Medical Expenditure Panel Survey data set}
\includegraphics[height=0.18\textheight]{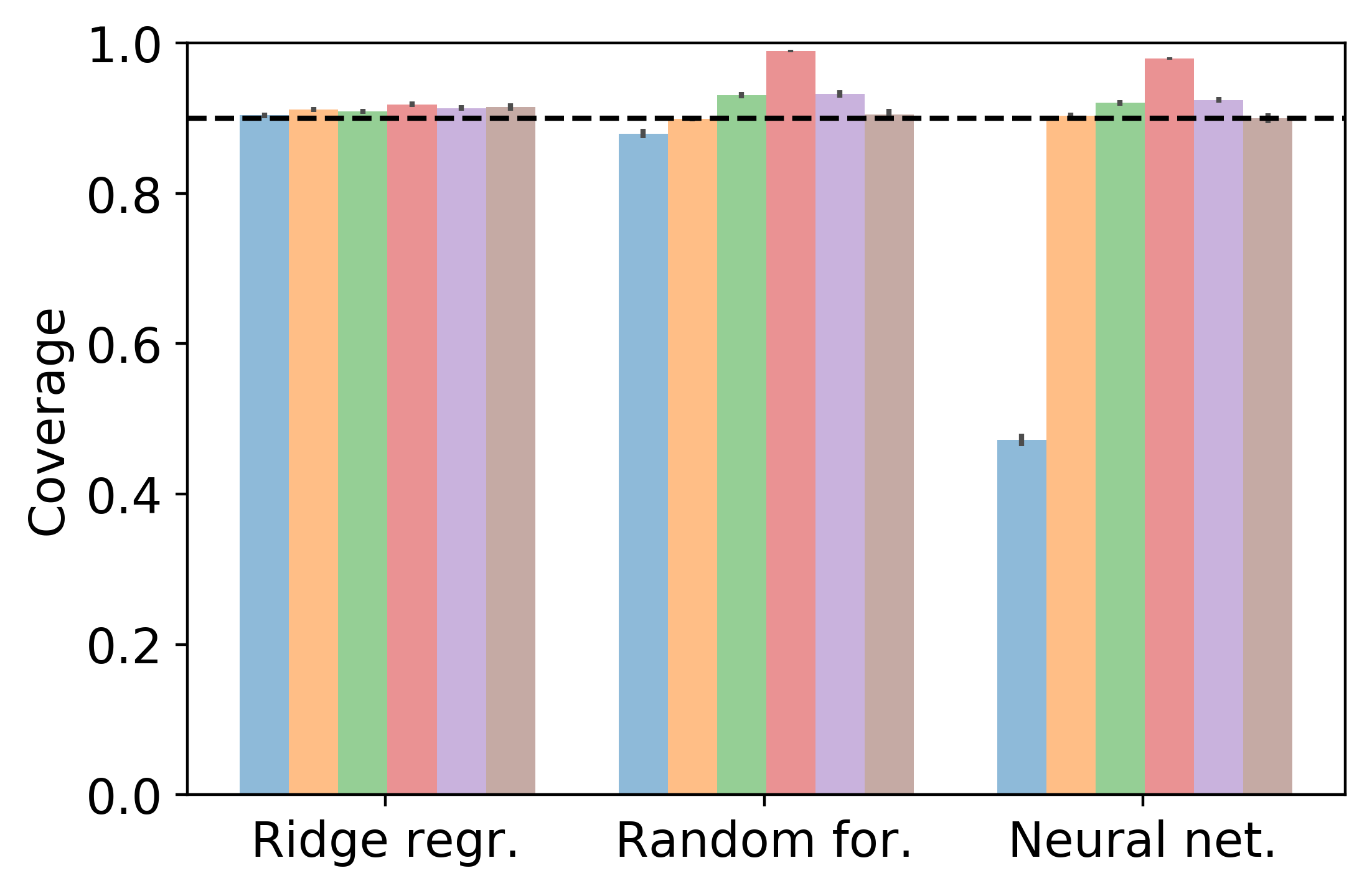}\quad 
\includegraphics[height=0.18\textheight]{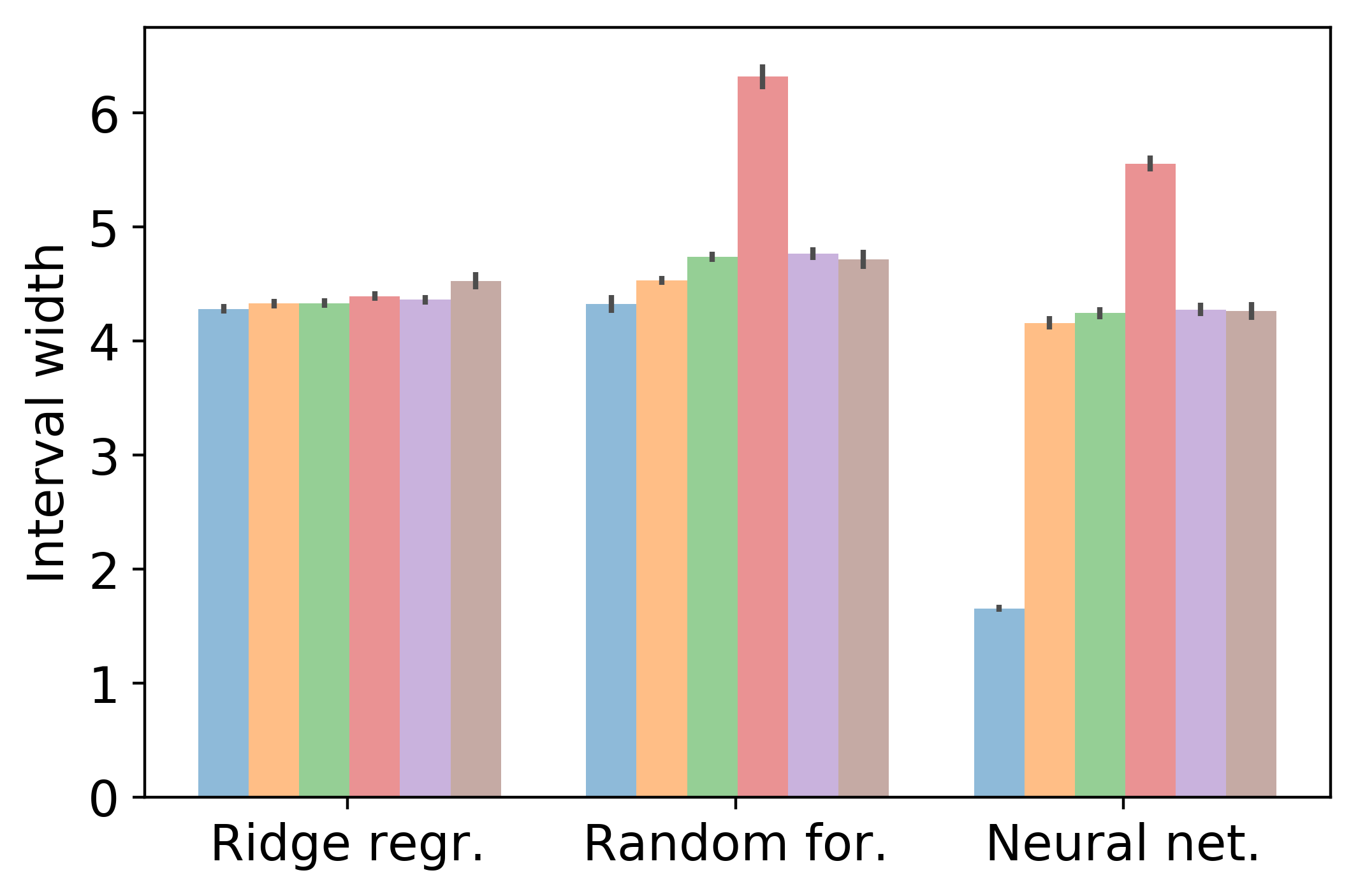}
\label{fig:realdata_results_meps}
\end{subfigure}
\caption{Results on three real data sets, using either ridge
  regression, random forests, or neural networks as the regression
  algorithm. The bar plots show the coverage and the width of the
  predictive interval for all methods. The figures display the mean
  over 20 independent trials (i.e., splits into training and test
  data), with error bars to show $\pm$ one standard
  error. In general, the naive method
    undercovers while jackknife-minmax overcovers, and the remaining
    methods have well calibrated coverage. In terms of their interval
    lengths, we typically (but not necessarily) get the expected
    order: jackknife $<$ jackknife+ $<$ 10-fold CV+ $<$ split
    conformal.}
\label{fig:realdata_results}
\end{figure}

\subsubsection{Methods}

Our procedure is the same for each of the three data sets. We randomly sample $n=200$ data points
from the full data set, to use as the training data. The remaining points form the test set.

We run our experiment using three different regression algorithms $\alg$---namely,
ridge regression, random forests, and neural networks.
The details of these algorithms is as follows:
\begin{itemize}
\item For ridge regression, we define
$\muh(x) =\widehat{\beta}_0 +  x^\top \widehat{\beta}$ for \[
\widehat{\beta}_0,\widehat{\beta}=\arg\min_{\beta_0\in\R,\beta\in\R^d}
\left\{\frac{1}{2}\sum_{i=1}^n (Y_i - \beta_0 - X_i^\top\beta)^2 + \lambda\big\|\beta\big\|^2_2\right\},\]
where the penalty parameter is chosen as $\lambda = 0.001\|X_{\textnormal{mat}}\|^2$, where $X_{\textnormal{mat}}\in\R^{n\times d}$ is the
covariate matrix of the training data, and $\|X_{\textnormal{mat}}\|$ is its spectral
norm.  This choice is to accommodate situations in which
  the matrix $X_{\textnormal{mat}}$ does not have full column rank as in the case where
  $d > n$. In such cases, the solution above is nearly the
  least-squares solution with minimum $\ell_2$ norm.
\item For random forests, we use the \texttt{RandomForestRegressor} method from
 the \texttt{scikit-learn} package~\citep{scikit-learn} in Python,
with 20 trees grown for each random forest using the mean absolute error criterion, and with default
settings otherwise.
\item For neural networks, we use the \texttt{MLPRegressor} method
  also from \texttt{scikit-learn}, run with the L-BFGS solver and the
  logistic activation function, and with default settings otherwise.
\end{itemize}
For each choice of $\alg$, we construct six prediction intervals (naive, jackknife, jackknife+, jackknife-minmax, CV+, split conformal), and calculate their empirical coverage
rate and their average width on the test set. We then repeat this procedure 20 times, with the train/test split formed randomly each time,
and report the mean and standard error over these 20 trials.

\subsubsection{Results}
Figure~\ref{fig:realdata_results} displays the results of the real data experiments. 
For each data set, each regression algorithm, and each one of the six prediction interval methods,
the figure plots
the average coverage and average width, together with their standard errors
 across the 20 independent trials.

 We see that the jackknife and jackknife+ methods both yield empirical
 coverage extremely close to the target level of 90\%, and have very
 similar predictive interval widths. However, in some settings, the
 jackknife+ shows slightly higher coverage than jackknife, and
 slightly wider prediction intervals.  These settings correspond to
 regression methods with greater instability.  
 As expected,
 the naive method undercovers in some settings and the
 jackknife-minmax is generally overly conservative. 
 Split conformal performs reasonably well: its length and coverage is sometimes comparable to the jackknife+, but is also significantly wider in some instances. Intuitively, if the best regression function in the considered function class is simple and the dataset is large, split conformal should perform fine even though it uses $n/2$ points for training and $n/2$ for calibration; however, in settings where the dataset is small relative to the complexity of the best regressor, then we should observe significant gains in using $n-1$ points for training and $n$ points for calibration. One phenomenon that is not visible in the empirical results is that split conformal is a randomized method, 
 with output varying slightly depending on the random split, while jackknife+ is a deterministic method on any fixed training data set.

 \section{Summary}\label{sec:discussion}

 The jackknife+ differs from the jackknife in that it uses the quantiles
 of
 \[
   \muh_{-i}(X_{n+1}) \pm R^{\textnormal{LOO}}_i = \muh(X_{n+1}) +\Big( \muh_{-i}(X_{n+1}) - \muh(X_{n+1})\Big)  \pm R^{\textnormal{LOO}}_i,
 \]
 instead of those of $\muh(X_{n+1}) \pm R^{\textnormal{LOO}}_i$, to
 build predictive intervals. By applying the shifts
 $\muh_{-i}(X_{n+1}) - \muh(X_{n+1})$, the jackknife+ effectively
 accounts for the (possible) algorithm instability, yielding rigorous
 coverage guarantees under no assumptions other than exchangeable
 samples. This, together with its empirical performance on real data,
 makes it a better choice than the jackknife in practice. In
 cases where the jackknife+ is computationally prohibitive, $K$-fold CV+ offers
 an attractive alternative.  Here, it would be interesting to see if the
 coverage guarantees for the latter method can be somewhat sharpened.

\section*{Acknowledgements}
The authors are grateful to the American Institute of Mathematics for
supporting and hosting our collaboration.  R.F.B.~was partially
supported by the National Science Foundation via grant DMS--1654076
and by an Alfred P.~Sloan fellowship.  E.J.C.~was partially supported
by the Office of Naval Research under grant N00014-16-1-2712, by the
National Science Foundation via grant DMS--1712800, and by a generous
gift from TwoSigma. The authors are grateful to an anonymous reviewer
for helpful suggestions on the presentation of the proof of Theorem~\ref{thm:main}. E.J.C.~thanks Yaniv Romano for help with some experiments. A.R.~thanks Arun Kumar Kuchibhotla for discussions regarding cross-conformal prediction.

\bibliographystyle{plainnat}
\bibliography{refs}

\appendix

\section{Asymmetric jackknife+ and CV+}\label{sec:extensions}
In settings where the distribution of $Y$ given $X$ appears to have symmetric noise, it is natural to construct predictions intervals
symmetrically, which is why we can consider absolute values of residuals. If the data is likely to be skewed,
however, we may want to consider an asymmetric
construction. Fix any $\alpha_{+},\alpha_{-}>0$ with $\alpha_{+} + \alpha_{-} = \alpha$,  and let
\begin{multline}\label{eqn:jackknife+_asym}
\Ch^{\textnormal{jackknife+}}_{n,\alpha_\pm}(X_{n+1}) ={}\\ 
\left[\qhm_{n,\alpha_{-}}\big\{\muh_{-i}(X_{n+1})  + R^{\textnormal{sgn,LOO}}_i\big\}, \   
\qhp_{n,\alpha_{+}}\big\{\muh_{-i}(X_{n+1}) + R^{\textnormal{sgn,LOO}}_i\big\}\right],\end{multline}
where the signed residuals are
\[R^{\textnormal{sgn,LOO}}_i = Y_i - \muh_{-i}(X_i).\]
We can of course consider the analogous asymmetric version of the original jackknife,
\begin{multline}\label{eqn:jackknife_asym}
\Ch^{\textnormal{jackknife}}_{n,\alpha_\pm}(X_{n+1}) ={}\\ 
\left[\muh(X_{n+1}) + \qhm_{n,\alpha_{-}}\big\{R^{\textnormal{sgn,LOO}}_i\big\}, \   
\muh(X_{n+1}) + \qhp_{n,\alpha_{+}}\big\{R^{\textnormal{sgn,LOO}}_i\big\}\right].\end{multline}
This type of asymmetric jackknife was considered by~\citet{steinberger2018conditional}.
Similarly we can define an asymmetric version of jackknife-minmax or of CV+.
We remark that, even if we were to choose $\alpha_{-}=\alpha_{+}=\alpha/2$, these asymmetric constructions would not necessarily be equal to the original jackknife, jackknife+, jackknife-minmax, and CV+ intervals, because the empirical distribution of the signed residuals will in general be asymmetric even if only due to random chance.

All of the coverage guarantees that we have proved for the various symmetric methods, hold
also for their asymmetric counterparts. For example, to verify $1-2\alpha$ coverage for the asymmetric jackknife+ in the assumption-free setting,
the proof of Theorem~\ref{thm:main} proceeds identically except that the matrix of residuals $R\in\R^{(n+1)\times(n+1)}$ constructed in the proof is replaced with two matrices
\[(R_\pm)_{ij} = \begin{cases} + \infty, & i=j,\\
      \pm \, (Y_i - \mut_{-(i,j)}(X_i)),& i\neq j,\end{cases}
  \]
  where $R_+$ (resp.~$R_{-}$) is used to bound the probability of
  noncoverage in the right (resp.~left) tail by $\alpha_{+}$
  (resp.~$\alpha_{-}$).

\section{Additional proofs}\label{sec:appendix_proofs}

\subsection{Proof of jackknife-minmax (Theorem~\ref{thm:minmax})}
The proof for jackknife-minmax proceeds nearly identically to the proof for jackknife+ (Theorem~\ref{thm:main}).
We define the residuals $R_{ij}$ exactly
as in the proof of Theorem~\ref{thm:main}, but we will use a different definition for the matrix $A$:
\[A_{ij} = \One{\min_{j'} R_{ij'} > R_{ji}},\]
that is, the {\em smallest} residual for data point $i$ (leaving out any point $j'$) is larger than $R_{ji}$, which is data point $j$'s residual 
when leaving out point $i$. Define $\mathcal{S}(A)$ exactly as before.
Now we follow essentially the same three steps as in the proof of Theorem~\ref{thm:main}:
\begin{itemize}
\item Step 1: we will establish deterministically that, for this new definition of the matrix $A$,
we have $|\mathcal{S}(A)|\leq \alpha(n+1)$ (whereas, for jackknife+, the bound was $2\alpha(n+1)$).
\item Step 2: using the fact that the data points are i.i.d.~(or more generally exchangeable), we will show that the probability that
the test point $n+1$ is strange (i.e., $n+1\in\mathcal{S}(A)$) is therefore bounded by $\alpha$.

\item Step 3: finally, we will verify that the jackknife-minmax interval can only fail to cover
the test response value $Y_{n+1}$ if $n+1$ is a strange point.
\end{itemize}

\paragraph{Step 1: bounding the number of strange points}
To prove Step 1, let
\[i_\star \in \arg\min_{i\in\mathcal{S}(A)} \min_{j'} R_{i,j'}.\]
Then by definition, for all $j\in \mathcal{S}(A)$,
$R_{ji_\star} \geq \min_{j'} R_{jj'} \geq \min_{j'}R_{i_\star j'}$.
This means that, by definition of the new comparison matrix $A$, we have $A_{i_\star j} = 0$ for all $j\in\mathcal{S}(A)$,
and therefore,
\[n+1 - |\mathcal{S}(A)|\geq \sum_{j=1}^{n+1} A_{i_\star j}\geq (1-\alpha)(n+1),\]
where the last step holds by definition of $\mathcal{S}(A)$ (since $i_\star\in\mathcal{S}(A)$ is a strange point).
Therefore $|\mathcal{S}(A)|\leq \alpha(n+1)$ as desired.

\paragraph{Step 2: exchangeability of the data points}
This step is identical to Step 2 in the proof of Theorem~\ref{thm:main}.

\paragraph{Step 3: connecting to jackknife-minmax}
Suppose that $Y_{n+1}\not\in\Ch^{\textnormal{jack-mm}}_{n,\alpha}(X_{n+1})$. 
This means that either
\[Y_{n+1} > \max_{i=1,\dots,n}\muh_{-i}(X_{n+1}) + \qhp_{n,\alpha}\big\{R^{\textnormal{LOO}}_j\big\},\]
which implies that $Y_{n+1} > \max_{i=1,\dots,n}\muh_{-i}(X_{n+1}) + R^{\textnormal{LOO}}_j$ for at least $(1-\alpha)(n+1)$ many indices $j\in\{1,\dots,n\}$,
or otherwise
\[Y_{n+1} < \min_{i=1,\dots,n}\muh_{-i}(X_{n+1}) -\qhp_{n,\alpha}\big\{R^{\textnormal{LOO}}_j\big\},\]
which implies that $Y_{n+1} < \min_{i=1,\dots,n}\muh_{-i}(X_{n+1}) - R^{\textnormal{LOO}}_j$ for at least $(1-\alpha)(n+1)$ many indices $j\in\{1,\dots,n\}$.
In either case, then, we have
\begin{align*}
(1-\alpha)(n+1)
 &\leq \sum_{j=1}^n \One{Y_{n+1} \not \in \big[\min_{i=1,\dots,n}\muh_{-i}(X_{n+1}) - R^{\textnormal{LOO}}_j, \max_{i=1,\dots,n}\muh_{-i}(X_{n+1}) + R^{\textnormal{LOO}}_j\big]}\\
&= \sum_{j=1}^n \One{ \min_{i=1,\dots,n}\big|Y_{n+1} - \muh_{-i}(X_{n+1})\big|>\big|Y_j - \muh_{-j}(X_j)\big| } \\
&= \sum_{j=1}^{n+1} \One{\min_{j'} R_{n+1,j'} >R_{j,n+1}} \\
&= \sum_{j=1}^{n+1} A_{n+1,j'},
\end{align*}
and therefore $n+1\in\mathcal{S}(A)$.

\subsection{Proofs for the CV+ method}

In this section, we will give details for how the CV+ method relates to the cross-conformal prediction method,
and then prove our theoretical guarantees for CV+.

\subsubsection{Details for comparing to the cross-conformal method}\label{sec:crossconf_app}
In Section~\ref{sec:crossconf}, we introduced the cross-conformal method~\eqref{eqn:crossconf} of \citet{vovk2015cross,vovk2018cross},
and stated two properties---first, that the cross-conformal prediction set is always contained in the CV+ prediction interval~\eqref{eqn:crossconf_vs_CV+},
and second, that  the results of \citet{vovk2018cross} imply a coverage guarantee~\eqref{eqn:crossconf_cov} for the cross-conformal method.
Here we give details to justify these two statements.

First, we verify that the cross-conformal prediction set $\Ch^{\textnormal{cross-conf}}_{n,K,\alpha}(X_{n+1})$ is contained in the CV+ interval.
To see this, suppose that $y\in \Ch^{\textnormal{cross-conf}}_{n,K,\alpha}(X_{n+1})$. Then by definition of this predictive set, we have
\[\tau + {\displaystyle \sum_{i=1}^n} \One{\big| y - \muh_{-S_{k(i)}}(X_{n+1})\big| < R^{\textnormal{CV}}_i} +\tau\One{\big| y - \muh_{-S_{k(i)}}(X_{n+1})\big| = R^{\textnormal{CV}}_i}  > \alpha (n+1)\]
for some $\tau\in[0,1]$. Since the left-hand side is monotone in $\tau$, in particular this implies that the above inequality holds at $\tau=1$, and so
\[ \sum_{i=1}^n \One{\big| y - \muh_{-S_{k(i)}}(X_{n+1})\big| \leq R^{\textnormal{CV}}_i} >\alpha(n+1) - 1.\]
Therefore,
\[ \sum_{i=1}^n \One{ y >  \muh_{-S_{k(i)}}(X_{n+1})+ R^{\textnormal{CV}}_i}  <  (1-\alpha)(n+1),\]
meaning that $y$ is {\em not} larger than the $\lceil(1-\alpha)(n+1)\rceil$-th smallest value of $\muh_{-S_{k(i)}}(X_{n+1})+ R^{\textnormal{CV}}_i$, $i=1,\dots,n$.
In other words,
\[y \leq \qhp_{n,\alpha}\big\{\muh_{-S_{k(i)}}(X_{n+1}) + R^{\textnormal{CV}}_i\big\}.\]
An identical argument proves that
\[y \geq \qhm_{n,\alpha}\big\{\muh_{-S_{k(i)}}(X_{n+1}) - R^{\textnormal{CV}}_i\big\},\]
meaning that $y$ must lie in the CV+ prediction interval $\Ch^{\textnormal{CV+}}_{n,K,\alpha}(X_{n+1})$.
This completes our verification of the claim~\eqref{eqn:crossconf_vs_CV+} that the CV+ interval contains the cross-conformal prediction set deterministically.

Next we give details for the coverage guarantee for the cross-conformal method, which is implied but not stated explicitly by \citet{vovk2018cross}.
Specifically, \citet{vovk2018cross} show that a modification of the $K$-fold cross-conformal method can lead to a $1-2\alpha$ coverage guarantee.
To define the modified method, let
\[P_k(y) = \frac{\tau + \sum_{i\in S_k} \One{ \big|y - \muh_{-S_k}(X_{n+1})\big| < R^{\textnormal{CV}}_i } + \tau\One{\big|y - \muh_{-S_k}(X_{n+1})\big| = R^{\textnormal{CV}}_i }}{m+1},\]
where $m=n/K$ is the number of data points in each fold (assumed to be an integer).
Plugging in the true test value $Y_{n+1}$, we see that $P_k(Y_{n+1})$ is
 a rank-based $p$-value comparing the test residual $\big|Y_{n+1}-\muh_{-S_k}(X_{n+1})\big|$ of the test point against all other residuals in the $k$th fold.
(Here $\tau\sim\textnormal{Unif}[0,1]$ corrects for discretization, so that this p-value is uniformly distributed on $[0,1]$ instead
of on a discrete grid.)
\citet{vovk2018cross} then consider a modified cross-conformal method,
\begin{equation}\label{eqn:mod_crossconf}\Ch^{\textnormal{modified-cc}}_{n,K,\alpha}(X_{n+1}) = \bigg\{y\in \R : \frac{1}{K}\sum_{k=1}^K P_k(y)>\alpha\bigg\}.\end{equation}
\citet{vovk2018cross} cite earlier work by \citet[Corollary 2]{vovk2012combining}, which proves that an arithmetic means of $p$-values is itself a valid p-value up to a factor of 2---that is,
\[\PP{\frac{1}{K}\sum_{k=1}^K P_k(Y_{n+1}) \leq \alpha} \leq 2\alpha\]
for any $\alpha\in[0,1]$, and so the modified cross-conformal method at level $\alpha$ has predictive coverage at least $1-2\alpha$.

Now we relate the modified cross-conformal method to its original version. Plugging in the definition of the p-values $P_k(y)$
and comparing with the original (unmodified) cross-conformal predictive set~\eqref{eqn:crossconf}, we can see that, deterministically,
\[\Ch^{\textnormal{cross-conf}}_{n,K,\alpha}(X_{n+1})\supseteq \Ch^{\textnormal{modified-cc}}_{n,K,\alpha'}(X_{n+1})\text{ where }\alpha' = \alpha + (1-\alpha)\frac{K-1}{n+K}.\]
Since the modified cross-conformal method, run at level $\alpha'$, has coverage at least $1-2\alpha'$, this proves that
\[\PP{Y_{n+1} \in \Ch^{\textnormal{cross-conf}}_{n,K,\alpha}(X_{n+1})} \geq 1- 2\alpha' \geq 1 - 2\alpha - 2(1-\alpha)\cdot \frac{1-1/K}{n/K+1}.\]
\subsubsection{Proof of Theorem~\ref{thm:CV+}} This proof
  follows essentially the same steps as the proof of
  Theorem~\ref{thm:main}.  Suppose that we draw $n/K-1$ additional
  test points, so that in total we have $m=n/K$ many test points,
  $(X_{n+1},Y_{n+1}),\dots$, $(X_{n+m},Y_{n+m})$. After partitioning
  the training data into sets $S_1,\dots,S_K$ of size $m$, we define
  $S_{K+1} = \{n+1,\dots,n+m\}$, the set of test points.
 
For any $k,k'\in\{1,\dots,K+1\}$ with $k\neq k'$, let $\mut_{-(S_k,S_{k'})}$ define the regression function fitted on the training plus test data, with subsets $S_k$ and $S_{k'}$ removed.  
Next, we define a matrix of residuals, $R\in\R^{(n+m)\times(n+m)}$ with entries
\[R_{ij} = \begin{cases} + \infty, & k(i)=k(j),\\
\big|Y_i - \mut_{-(S_{k(i)},S_{k(j)})}(X_i)\big|,& k(i)\neq k(j).\end{cases}\]
Define a comparison matrix $A\in\{0,1\}^{(n+m)\times(n+m)}$ with
\[A_{ij} = \One{R_{ij}>R_{ji}},\]
and consider the set of ``strange'' points,
\[\mathcal{S}(A) = \left\{ i\in\{1,\dots,n+m\} : A_{i\bullet} \geq (1-\alpha)(n+1)\right\},\]
where $A_{i\bullet}=\sum_{j=1}^{n+m} A_{ij}$.
Now we proceed as for the proof of Theorem~\ref{thm:main}.\begin{itemize}
\item
First, for Step 1, we bound the number of strange points deterministically as $|\mathcal{S}(A)|\leq 2\alpha (n+m) + (1-2\alpha)(m-1)-1$.
\item For Step 2 we see that the exchangeability of the data points implies that  the probability that the test point $n+1$ is strange (i.e., $n + 1 \in\mathcal{S}(A)$)
 is therefore bounded by $2\alpha + \frac{1-K/n}{K+1}$. 
 \item For Step 3,  we see that noncoverage of the CV+ interval implies that the test point is strange, which completes the proof of the theorem.
\end{itemize}
For Step 3, the proof that $Y_{n+1}\not\in  \Ch^{\textnormal{CV+}}_{n,K,\alpha}(X_{n+1})$ implies that $n+1\in\mathcal{S}(A)$,
is identical to the corresponding step in Theorem~\ref{thm:main}. 
We remark that in fact, our argument verifies a strictly stronger statement: if $Y_{n+1}$ is not contained in the $K$-fold cross-conformal 
prediction set (with $n=K$, in the case of jackknife+), 
then the test point is strange---this is strictly stronger because the cross-conformal prediction set satisfies $
\Ch^{\textnormal{cross-conf}}_{n,K,\alpha}(X_{n+1})\subseteq \Ch^{\textnormal{CV+}}_{n,K,\alpha}(X_{n+1})$ always.

Turning to Step 1, we now bound the number of strange points. This step is very similar to the proof of Step 1
in the proof of Theorem~\ref{thm:main}.
Let $s_k = |S_k \cap \mathcal{S}(A)|$ be the number of strange points
in the $k$th fold, so that $s_1+\dots+s_{K+1}=s := |\mathcal{S}(A)|$.
As before, each strange point $i \in \mathcal{S}(A)$ can lose
against at most $\alpha(n+1)-1$ other strange points. However,
the difference relative to jackknife+ is that the data points $i$ and $j$ in the same fold ($k(i)=k(j)$)
do not play against each other in the ``tournament'' so there are two types of pairs of strange points:
those that are in different folds (writing $s=|\mathcal{S}(A)|$, there are at most $s\cdot \big(\alpha(n+1)-1\big)$
such pairs, as in the jackknife+ proof), and those that are in the same fold (and therefore do not play a game).
 Thus we have established
that
\[ \frac{s(s-1)}{2} \leq s \cdot \big(\alpha(n+1)-1\big) + \sum_k \frac{s_k(s_k-1)}{2}.\]
We can simplify  the last sum as
\[\sum_k \frac{s_k(s_k-1)}{2} =\sum_k \frac{s_k^2-s_k}{2}\leq \sum_k \frac{m s_k-s_k}{2} = \frac{m-1}{2}\sum_k s_k = \frac{s(m-1)}{2},\]
since $s_k \leq |S_k| = n/K= m$ for each $k$.
Simplifying the expression above we have therefore proved that
\[ s\leq  2\alpha(n+1) + m - 2 = 2\alpha(n+m) + (1-2\alpha)(m-1) - 1,\]
as desired.

Finally, we verify Step 2. For the $K$-fold setting, this is a bit more subtle than for jackknife+. This is because we cannot claim that $A \eqd \Pi A \Pi^\top$ for
 any $(n+m)\times(n+m)$ permutation matrix $\Pi$---indices $i,j$ belonging in the same fold ($k(i)=k(j)$)
behave differently than indices $i,j$ in different folds ($k(i)\neq k(j)$). However,
treating the split of the training and test data into folds $S_1,\dots,S_K,S_{K+1}$ as fixed,
we can verify that $A \eqd \Pi A \Pi^\top$ for any $(n+m)\times(n+m)$ permutation matrix $\Pi$ that preserves the equivalence relation induced
by the folds, $i\sim j$ if $k(i)=k(j)$. Now, for any $j\in\{1,\dots,n+m\}$, there exists such a $\Pi$ with $\Pi_{j,n+1}=1$.
Thus, as in the proof of Theorem~\ref{thm:main}, this implies that $\PP{n+1 \in\mathcal{I}(A)}  = \PP{j\in\mathcal{I}(A)} $ for all $j\in\{1,\dots,n+m\}$.
Therefore, combining with the result of Step 1,
\[\PP{n+1\in\mathcal{S}(A)} \leq \frac{2\alpha (n+m) + (1-2\alpha)(m-1)-1}{n+m} \leq 2\alpha + \frac{1-K/n}{K+1}.\]
Combining with Step 3, we have completed the proof of the coverage guarantee.

\subsection{Proof of Theorem~\ref{thm:out_of_sample_stability}}\label{sec:proof_thm:out_of_sample_stability}
We will first consider an oracle leave-one-out method that, while impossible to implement in practice, achieves
the target $1-\alpha$ coverage rate. We will then relate the $\eps$-inflated jackknife and $2\eps$-inflated jackknife+ to the oracle method.

\subsubsection{Oracle method}
For each $i=1,\dots,n+1$, let  $\mut_{-i}$ be the regression function
fitted on the training and test data with point $i$ removed, i.e.
\[\mut_{-i} = \alg\Big((X_1,Y_1),\dots,(X_{i-1},Y_{i-1}),(X_{i+1},Y_{i+1}),\dots,(X_{n+1},Y_{n+1})\Big).\]
Note that $\mut_{-(n+1)} = \muh$, while for $i=1,\dots,n$, $\mut_{-i}$ differs from $\muh_{-i}$ since the test point $(X_{n+1},Y_{n+1})$ is included in the regression.
Consider an ``oracle jackknife'' method, where we use $\mut_{-i}$ in place of $\muh_{-i}$:
\[\Ch^{\textnormal{oracle}}_{n,\alpha'}(X_{n+1}) = \muh(X_{n+1}) \pm \qhp_{n,\alpha'}\big\{R^{\textnormal{oracle}}_i\big\},\]
where $\alpha' = \alpha + \sqrt{\nu} $ and
\[R^{\textnormal{oracle}}_i = \big|Y_i - \mut_{-i}(X_i)\big|.\]
Next, we will confirm that this oracle method achieves $1-\alpha'$ coverage. This fact is based on the exchangeability of 
the training and test data points, and can be proved using standard techniques from the conformal prediction literature (see, e.g., \citet{vovk2005algorithmic,lei2018distribution} for background).
Writing
\[R^{\textnormal{oracle}}_{n+1} =  \big|Y_{n+1} - \mut_{-(n+1)}(X_{n+1})\big|  =  \big|Y_{n+1} - \muh(X_{n+1})\big|,\] 
we see that failure to cover, i.e., $Y_{n+1}\not\in\Ch^{\textnormal{oracle}}_{n,\alpha'}$, occurs if and only if
\begin{equation}\label{eqn:oracle_step1}R^{\textnormal{oracle}}_{n+1} > \textnormal{the $\lceil (1-\alpha')(n+1)\rceil$-th smallest value of }R^{\textnormal{oracle}}_1,\dots,R^{\textnormal{oracle}}_n.\end{equation}
Now, since the training and test data are i.i.d., and the algorithm $\alg$ is assumed to be invariant to the labeling of the points~\eqref{eqn:alg_invariant_reordering}, this means that the resulting oracle residuals $R^{\textnormal{oracle}}_1,\dots,R^{\textnormal{oracle}}_n,R^{\textnormal{oracle}}_{n+1}$ are exchangeable. In other words, the rank of $R^{\textnormal{oracle}}_{n+1}$ among this list of oracle residuals is uniformly random. Therefore, the probability that the event in~\eqref{eqn:oracle_step1} occurs is equal to the probability that $R^{\textnormal{oracle}}_{n+1}$ is not one of the smallest $\lceil (1-\alpha')(n+1)\rceil$ values in the list of oracle residuals, and so
\begin{equation}\label{eqn:oracle_coverage}\PP{Y_{n+1}\not\in\Ch^{\textnormal{oracle}}_{n,\alpha'}(X_{n+1}) }\leq 1 - \frac{\lceil (1-\alpha')(n+1)\rceil}{n+1} \leq \alpha'.\end{equation}
(The first inequality cannot be replaced with an equality, due to the possibility of ties among the residuals.)

\subsubsection{Bound for the jackknife}
Now we relate the oracle method back to the jackknife. 
We will show that
\[\Ch^{\textnormal{jackknife},\eps}_{n,\alpha}(X_{n+1})\supseteq \Ch^{\textnormal{oracle}}_{n,\alpha'}(X_{n+1}) \]
with sufficiently high probability. To see why, suppose instead that this set inclusion does not hold. Then  by definition of $\Ch^{\textnormal{jackknife},\eps}_{n,\alpha}(X_{n+1})$, we must have
\[ \qhp_{n,\alpha}\big\{R^{\textnormal{LOO}}_i\big\} + \eps < \qhp_{n,\alpha'}\big\{R^{\textnormal{oracle}}_i\big\}.\]
By definition of these quantiles, we can conclude that the number of indices $i\in\{1,\dots,n\}$ with $R^{\textnormal{oracle}}_i > R^{\textnormal{LOO}}_i + \eps$
is at least
\[\lceil (1-\alpha)(n+1)\rceil - \Big(\lceil (1-\alpha')(n+1)\rceil - 1\Big) \geq \sqrt{\nu} (n+1).\]
Therefore,
\begin{multline}
\PP{\Ch^{\textnormal{jackknife},\eps}_{n,\alpha}(X_{n+1})\not\supseteq \Ch^{\textnormal{oracle}}_{n,\alpha'}(X_{n+1})}\\
\leq \PP{\sum_{i=1}^n \One{R^{\textnormal{oracle}}_i > R^{\textnormal{LOO}}_i + \eps} \geq \sqrt{\nu} (n+1) }\\
\leq \PP{\sum_{i=1}^n \One{\big|\mut_{-i}(X_i) - \muh_{-i}(X_i)\big|> \eps} \geq \sqrt{\nu} (n+1) }\\
\label{eqn:oracle_in_jackknife}\leq\frac{\EE{\sum_{i=1}^n \One{\big|\mut_{-i}(X_i) - \muh_{-i}(X_i)\big| >  \eps}}}{\sqrt{\nu} (n+1) },
\end{multline}
where the last step holds by Markov's inequality.
Observe also that, for every $i=1,\dots,n$,
\begin{align*}
\notag\PP{\big|\mut_{-i}(X_i) - \muh_{-i}(X_i)\big| >  \eps} &=\PP{\big|\mut_{-i}(X_i) - \mut_{-(i,n+1)}(X_i)\big| >  \eps} \\
\notag&=\PP{\big|\mut_{-(n+1)}(X_{n+1}) - \mut_{-(n+1,i)}(X_{n+1})\big| >  \eps} \\
\notag&= \PP{\left| \muh(X_{n+1}) - \muh_{-i}(X_{n+1})\right| > \eps}\\
& \leq \nu,\end{align*}
where the first and third step hold by definition of the various regression functions; the 
second step holds since the data points are i.i.d.~and so swapping the label
of point $i$ and point $n+1$ does not change the distribution; and the last step holds by applying out-of-sample stability~\eqref{eqn:out_of_sample_stability}.
Therefore, returning to~\eqref{eqn:oracle_in_jackknife}, we have
\[\PP{\Ch^{\textnormal{jackknife},\eps}_{n,\alpha}(X_{n+1})\not\supseteq \Ch^{\textnormal{oracle}}_{n,\alpha'}(X_{n+1})} \leq \frac{n \cdot \nu }{\sqrt{\nu} (n+1)} \leq \sqrt{\nu}.\]
Combining this bound with~\eqref{eqn:oracle_coverage}, we have therefore proved that
\begin{multline}\label{eqn:proof_eps_jackknife}\PP{Y_{n+1}\in\Ch^{\textnormal{jackknife},\eps}_{n,\alpha}(X_{n+1})} \geq \PP{Y_{n+1}\not\in\Ch^{\textnormal{oracle}}_{n,\alpha'}(X_{n+1}) }  - \sqrt{\nu}\\\geq 1 - \alpha' - \sqrt{\nu} = 1-\alpha - 2\sqrt{\nu}.\end{multline}

\subsubsection{Bound for the jackknife+}
The argument for the jackknife+ proceeds similarly, except that we now compare against the jackknife rather than the oracle---we will 
verify that
\[\Ch^{\textnormal{jackknife+},2\eps}_{n,\alpha}(X_{n+1})\supseteq \Ch^{\textnormal{jackknife},\eps}_{n,\alpha'}(X_{n+1}) \]
holds
with sufficiently high probability, where we again define $\alpha' = \alpha + \sqrt{\nu}$.

Suppose that this does not hold. Then it must either fail at the upper bounds of the intervals, i.e.,
\[\qhp_{n,\alpha}\big\{\muh_{-i}(X_{n+1}) + R^{\textnormal{LOO}}_i\big\}  + 2\eps <\muh(X_{n+1}) +  \qhp_{n,\alpha'}\big\{R^{\textnormal{LOO}}_i\big\} + \eps\]
or at the lower bounds, i.e.,
\[\qhm_{n,\alpha}\big\{\muh_{-i}(X_{n+1}) -  R^{\textnormal{LOO}}_i\big\}  -  2\eps  >\muh(X_{n+1})   - \qhp_{n,\alpha'}\big\{ R^{\textnormal{LOO}}_i\big\} - \eps.\]
In the first case, this implies that $\muh(X_{n+1}) > \muh_{-i}(X_{n+1}) + \eps$ for at least $\sqrt{\nu} (n+1) $ many indices $i\in\{1,\dots,n\}$,
while in the second case, we instead have $\muh(X_{n+1}) < \muh_{-i}(X_{n+1}) - \eps$ for at least $\sqrt{\nu} (n+1) $ many indices $i\in\{1,\dots,n\}$.
Combining the two, then, we have
\begin{multline}
\PP{\Ch^{\textnormal{jackknife+},2\eps}_{n,\alpha}(X_{n+1})\not\supseteq \Ch^{\textnormal{jackknife},\eps}_{n,\alpha'}(X_{n+1})}\\
\leq \PP{\sum_{i=1}^n \One{\big|\muh(X_{n+1}) - \muh_{-i}(X_{n+1})\big|> \eps} \geq \sqrt{\nu} (n+1)}\\
\leq \frac{\EE{\sum_{i=1}^n \One{\big|\muh(X_{n+1}) - \muh_{-i}(X_{n+1})\big|>  \eps}}}{\sqrt{\nu} (n+1) }
\label{eqn:jackknife_in_jackknife+}\leq \frac{\nu n}{\sqrt{\nu} (n+1)} \leq \sqrt{\nu},
\end{multline}
where we again apply Markov's inequality and the out-of-sample stability property just as for the jackknife proof.
Combining the coverage result~\eqref{eqn:proof_eps_jackknife} (with $\alpha'$ in place of $\alpha$)
with~\eqref{eqn:jackknife_in_jackknife+}, we have proved that
\begin{multline*}\PP{Y_{n+1}\in\Ch^{\textnormal{jackknife+},2\eps}_{n,\alpha}(X_{n+1})} \geq \PP{Y_{n+1}\not\in\Ch^{\textnormal{jackknife},\eps}_{n,\alpha'}(X_{n+1}) }  - \sqrt{\nu}\\\geq 
\left[ 1-\alpha'- 2\sqrt{\nu} \right] - \sqrt{\nu} = 1-\alpha - 4\sqrt{\nu}.\end{multline*}

\subsection{Proof of Theorem~\ref{thm:in_sample_stability}}\label{sec:proof_thm:in_sample_stability}
Let $R^{\textnormal{naive}}_i = \big|Y_i - \muh(X_i)\big|$ be the $i$th residual in the ``naive'' method while $R^{\textnormal{LOO}}_i = \big|Y_i - \muh_{-i}(X_i)\big|$ is the leave-one-out residual as before. Suppose that we run jackknife at level $1-\alpha'$, where $\alpha' =  \alpha + \sqrt{\nu}$. 
Then by definition of the two methods, we have
\[\Ch^{\textnormal{naive},2\eps}_{n,\alpha}(X_{n+1}) = \muh(X_{n+1}) \pm \big(2\eps+ \qhp_{n,\alpha}\big\{R^{\textnormal{naive}}_i\big\}\big)\]
and
\[\Ch^{\textnormal{jackknife},\eps}_{n,\alpha'}(X_{n+1}) = \muh(X_{n+1}) \pm \big(\eps +\qhp_{n,\alpha'}\big\{R^{\textnormal{LOO}}_i\big\}\big).\]
We will now check that 
\[\Ch^{\textnormal{naive},2\eps}_{n,\alpha}(X_{n+1})\supseteq \Ch^{\textnormal{jackknife},\eps}_{n,\alpha'}(X_{n+1})\]
with sufficiently high probability. Following similar arguments as in the proof of Theorem~\ref{thm:out_of_sample_stability},
if this does not hold, then it must be the case that
\[\sum_{i=1}^n \One{R^{\textnormal{LOO}}_i > R^{\textnormal{naive}}_i + \eps} \geq \lceil (1-\alpha)(n+1) \rceil -  \Big(\lceil (1-\alpha')(n+1) \rceil  - 1\Big)\geq \sqrt{\nu} (n+1).\]
By the triangle inequality, this implies that
\[\sum_{i=1}^n \One{\big|\muh(X_i) - \muh_{-i}(X_i)\big|> \eps} \geq  \sqrt{\nu} (n+1).\]
Therefore, by Markov's inequality,
\begin{multline*}\PP{\Ch^{\textnormal{naive},2\eps}_{n,\alpha}(X_{n+1})\not\supseteq \Ch^{\textnormal{jackknife},\eps}_{n,\alpha'}(X_{n+1})} \\{}\leq
\frac{\EE{\sum_{i=1}^n \One{\big|\muh(X_i) - \muh_{-i}(X_i)\big|> \eps}}}{\sqrt{\nu} (n+1)}
\leq \frac{n \nu}{\sqrt{\nu} (n+1)} \leq \sqrt{\nu},
\end{multline*}
where the second step applies the in-sample stability property~\eqref{eqn:in_sample_stability}.
Therefore,
\begin{multline*}\PP{Y_{n+1}\in\Ch^{\textnormal{naive},2\eps}_{n,\alpha'}(X_{n+1})} \geq \PP{Y_{n+1}\in\Ch^{\textnormal{jackknife},\eps}_{n,\alpha'}(X_{n+1})} - \sqrt{\nu} \\
\geq\left[ 1-\alpha' - 2\sqrt{\nu} \right] - \sqrt{\nu}
\geq  1-\alpha - 4\sqrt{\nu} , \end{multline*}
where for the next-to-last step we apply the result of Theorem~\ref{thm:out_of_sample_stability} (with $\alpha'$ in place of $\alpha$).

\subsection{Proof of Theorem~\ref{thm:worstcase}}\label{sec:proof_thm:worstcase}
In this section, we will prove a stronger version of Theorem~\ref{thm:worstcase}, and will verify that 
the lower bounds on coverage hold even when we use the $\eps$-inflated versions of each of the intervals, for any $\eps > 0$.
 That is, we will construct pathological examples for which, for the naive method and for jackknife, we have
\[
\PP{Y_{n+1}\in\Ch^{\textnormal{naive},\eps}_{n,\alpha}(X_{n+1})}=\PP{Y_{n+1}\in\Ch^{\textnormal{jackknife},\eps}_{n,\alpha}(X_{n+1})}=0,
\]
and for jackknife+ with $\alpha\leq \frac{1}{2}$, we have
\[\PP{Y_{n+1}\in\Ch^{\textnormal{jackknife+},\eps}_{n,\alpha}(X_{n+1})}\leq 1 - 2\alpha + 6\sqrt{\frac{\log n}{n}}.\]

\subsubsection{Proof for jackknife and naive methods}
First we construct a simple example for the jackknife and naive intervals.
Define the regression method $\alg$ as follows: given a training sample of size $n$,
the algorithm returns the function 
\[\muh(x) = \begin{cases} 0,&\text{ if $x=X_i$ for any of the training points $X_i$, $i=1,\dots,n$,}\\
(1+\eps) n,&\text{ otherwise.}\end{cases}\] Now we define the data distribution on $(X,Y)$: let $X\sim \mathcal{N}(0,1)$,
and let $Y\equiv 0$.
Then with probability 1, the values $X_1,\dots,X_{n+1}$ will be distinct. In this case,
the leave-one-out residuals will be given by
\[R^{\textnormal{LOO}}_i = \big|Y_i - \muh_{-i}(X_i)\big| = \big|Y_i - (1+\eps)(n-1)\big| =  (1+\eps)(n-1)\]
for all $i=1,\dots,n$, yielding a residual quantile $\qhp_{n,\alpha}\big\{R^{\textnormal{LOO}}_i\big\} = (1+\eps)(n-1)$ and a prediction interval 
\[\Ch^{\textnormal{jackknife}}_{n,\alpha}(X_{n+1}) = \muh(X_{n+1}) \pm \qhp_{n,\alpha}\big\{R^{\textnormal{LOO}}_i\big\} = (1+\eps)n \pm (1+\eps)(n-1) = [1+\eps, (1+\eps)(2n-1)].\]
Therefore, the $\eps$-inflated interval is given by
\[\Ch^{\textnormal{jackknife},\eps}_{n,\alpha}(X_{n+1}) = [1, \eps + (1+\eps)(2n-1)].\]
However, $Y_{n+1}=0$ with probability 1, and so 
\[\PP{Y_{n+1}\in\Ch^{\textnormal{jackknife},\eps}_{n,\alpha}(X_{n+1})} = 0.\]
Similarly, for the naive method, its residuals will be given by
\[R^{\textnormal{naive}}_i = \big|Y_i - \muh(X_i)\big| = \big|Y_i - 0\big| =  0,\]
and so following the same argument we see that 
\[\PP{Y_{n+1}\in\Ch^{\textnormal{naive},\eps}_{n,\alpha}(X_{n+1})} = 0.\]
A simple calculation shows that in this example, jackknife+ interval contains 0 at its left endpoint, and hence maintains its coverage.

\subsubsection{Proof for jackknife+}
Next we give the construction for the jackknife+. Our construction is similar  in spirit to the example constructed by~\citet[Appendix A]{vovk2015cross}, which gives intuition for why 
the leave-one-out cross-conformal predictor (described earlier in Section~\ref{sec:jackknife+}) may fail when the $n+1$ data points are only assumed to be exchangeable
(specifically, the data points and their residuals are chosen deterministically, and then randomly permuted).
Here our construction is more technical as we need to work in the setting of i.i.d.~data.

To give intuition we first sketch the idea.
Suppose that the distribution of $(X,Y)$ is chosen such that, with probability $\approx 2\alpha$, $X$ is drawn from some ``bad'' region where predicting $Y$ is challenging
and we consistently underestimate $Y$, while with the remaining probability, $X$ is drawn from some ``good'' region where predicting $Y$ is easy---in fact, we can do this with zero error.
Now, what is the chance that $Y_{n+1}$ is not covered by the jackknife+? If $X_{n+1}$ is ``good'', then $Y_{n+1}$ will be covered. If $X_{n+1}$ is ``bad'', then we will have 
\begin{itemize}
\item For  approximately half of the ``bad'' $X_i$ ($\approx \alpha n$ data points), we will have
\[ 0 < Y_{n+1} - \muh_{-i}(X_{n+1}) < Y_i - \muh_{-i}(X_i).\]
\item For all ``good'' $X_i$ and for the remaining ``bad'' $X_i$ (in total, $\approx (1-\alpha)n$ of the data points),
we will have
\[Y_{n+1} - \muh_{-i}(X_{n+1}) >  Y_i - \muh_{-i}(X_i)\geq 0.\]
\end{itemize}
This will be sufficient to see that $Y_{n+1}$ is almost certainly not covered by the jackknife+ interval
whenever $X_{n+1}$ is ``bad'', i.e., with probability $2\alpha$.

Now we give the formal construction.
Fix a small $\gamma>0$ and a large $\tau >0$, which we will specify later on.
First, we will choose the distribution for the data. Let\footnote{To obtain this distribution, if the dimension is $d\geq 3$ we can take the above distribution over the first 3 coefficients in $X_i$ and simply ignore the remaining coefficients,
while if $d=1$ or $d=2$ we can transform the data if needed; for example if $d=1$, taking $X_i\sim\textnormal{Unif}[0,1]$, 
we can use the first $k$ digits of $X$ to generate a draw  $A_i\sim \textnormal{Bernoulli}(p)$ where $p\approx \big(2\alpha(1-\gamma)\big)$ (up to $10^{-k}$ accuracy),
then we use the next digit of $X_i$ to draw $B_i\sim  \textnormal{Unif}\{\pm 1\}$, and finally let $C_i\sim \textnormal{Unif}[-1,1]$ be defined by the remaining digits of $X_i$.}
\[X_i = (A_i,B_i,C_i)\sim \textnormal{Bernoulli}\big(2\alpha(1-\gamma)\big) \times \textnormal{Unif}\{\pm 1\}\times \textnormal{Unif}[-1,1],\]
and let
$Y_i = \tau A_i$.
Next we define the regression algorithm $\alg$ as follows: given a training sample $(X_j,Y_j) = \big((A_j,B_j,C_j),Y_j\big)$ indexed
over $j=1,\dots,m$, the resulting fitted regression function $\muh$ is defined as
\[\muh(x) = \tau  a c \cdot \prod_{j=1}^m B_j,\]
at any point $x=(a,b,c)\in\{0,1\}\times\{\pm 1\}\times[-1,1]$. Defining $B_{-i} = \prod_{j=1,\dots,n;j\neq i}B_j$, we therefore have
\[\muh_{-i}(x) = \tau  a  c \cdot B_{-i}\]
for each $i=1,\dots,n$.
Now we check that coverage is roughly $1-2\alpha$. We have
\begin{align*}
&\PP{Y_{n+1}\not\in\Ch_{n,\alpha}^{\textnormal{jackknife+},\eps}(X_{n+1})} \\
&\geq \PP{Y_{n+1} > \qhp_{n,\alpha}\big\{\muh_{-i}(X_{n+1}) + R^{\textnormal{LOO}}_i + \eps\big\}}\\
&\geq \PP{Y_{n+1} > \qhp_{n,\alpha}\big\{\muh_{-i}(X_{n+1}) + R^{\textnormal{LOO}}_i+ \eps\big\}\text{ and }A_{n+1}=1}\\
&= 2\alpha(1-\gamma)\cdot \PP{\tau > \qhp_{n,\alpha}\big\{ \tau C_{n+1} B_{-i} + R^{\textnormal{LOO}}_i + \eps\big\}} \\
&= 2\alpha(1-\gamma)\cdot \PP{\tau > \qhp_{n,\alpha}\big\{ \tau C_{n+1} B_{-i} + \tau A_i \big(1 - C_i B_{-i}\big) + \eps\big\}}\\
&= 2\alpha(1-\gamma)\cdot \PP{\qhp_{n,\alpha}\big\{  C_{n+1} B_{-i} +  A_i \big(1 - C_i B_{-i}\big) +\frac{\eps}{\tau}\big\}< 1}\\
&= 2\alpha(1-\gamma)\cdot\PP{\sum_{i=1}^n \One{ C_{n+1} B_{-i} +  A_i \big(1 - C_i B_{-i}\big) +\frac{\eps}{\tau}  < 1} \geq (1-\alpha)(n+1)}.
\end{align*}
Next we verify that this last probability is close to 1. 
These indicator variables are independent conditional on $A_1,\dots,A_n,B_1,\dots,B_n,C_{n+1}$ (as they then depend only on $C_i$ for each $i$).
We denote the conditional probabilities by
\[P_i = \PPst{C_{n+1} B_{-i} +  A_i \big(1 - C_i B_{-i}\big) +\frac{\eps}{\tau}  < 1}{A_1,\dots,A_n,B_1,\dots,B_n,C_{n+1}},\]
and calculate
\[P_i = \One{C_{n+1} B_{-i} < 1 - \eps/\tau}\cdot \begin{cases}
1, & \text{ if $A_i=0$,}\\
\frac{1 - B_{-i} C_{n+1} - \eps/\tau}{2}, & \text{ if $A_i=1$.}\\\end{cases}\]
By Hoeffding's inequality, we have
\begin{multline*}
\mathbb{P}\Bigg\{\sum_{i=1}^n \One{ C_{n+1} B_{-i} +  A_i \big(1 - C_i B_{-i}\big) +\frac{\eps}{\tau}  < 1}  \\{}\geq
 \sum_{i=1}^n P_i - t \  \Big\vert \ A_1,\dots,A_n,B_1,\dots,B_n,C_{n+1}\Bigg\} \geq 1- e^{-2t^2/n},\end{multline*}
for any $t\geq 0$. Choosing $t= \sqrt{\frac{n\log n}{2}}$ and marginalizing, we have
\[\PP{\sum_{i=1}^n \One{ C_{n+1} B_{-i} +  A_i \big(1 - C_i B_{-i}\big) +\frac{\eps}{\tau}  < 1} \geq \sum_{i=1}^n P_i - \sqrt{\frac{n\log n}{2}}} \geq 1- \frac{1}{n}.\]
Combining everything so far, we have therefore proved that
\begin{multline*}\PP{Y_{n+1}\not\in\Ch_{n,\alpha}^{\textnormal{jackknife+},\eps}(X_{n+1})}  \geq {}\\2\alpha(1-\gamma)\cdot \left(1 - \frac{1}{n}- \PP{\sum_{i=1}^n P_i - \sqrt{\frac{n\log n}{2}} < 
(1-\alpha)(n+1)}\right).\end{multline*}
Now we bound this last probability. We have
\begin{align*}
\sum_{i=1}^n P_i
 &= \sum_{i=1}^n \One{C_{n+1} B_{-i} < 1 - \eps/\tau}  \cdot \left( \One{A_i=0} + \One{A_i=1} \cdot \frac{1 - B_{-i}C_{n+1} - \eps/\tau}{2}\right)\\
&\geq \One{|C_{n+1}| \leq 1- \eps/\tau} \cdot  \sum_{i=1}^n  \left( (1-A_i) + A_i \cdot \frac{1 - B_{-i}C_{n+1} - \eps/\tau}{2}\right)\\
&\geq \One{|C_{n+1}| \leq 1- \eps/\tau} \cdot  \left( n - \frac{1+\eps/\tau}{2}\sum_{i=1}^n A_i\right) - \frac{1}{2}\left| \sum_{i=1}^n A_iB_i\right|,
\end{align*}
where the last step holds since $A_iB_{-i}C_{n+1} =A_iB_i \cdot  \big(C_{n+1}\prod_{j=1}^n B_j \big)$.
Next, $|C_{n+1}| \leq 1- \eps/\tau$ holds with probability $1-\eps/\tau$, while by Hoeffding's inequality,
\[\PP{\sum_{i=1}^n A_i \leq n\cdot 2\alpha(1-\gamma) + \sqrt{\frac{n\log n}{2}}}\geq 1-\frac{1}{n}\]
and
\[\PP{ \frac{1}{2}\left| \sum_{i=1}^n A_iB_i\right| \leq \sqrt{\frac{n\log n}{2}}}\geq 1 - \frac{2}{n}.\]
Putting these calculations together, 
\[\PP{\sum_{i=1}^n P_i \geq \left( n - \frac{1+\eps/\tau}{2}\left(n\cdot 2\alpha(1-\gamma) + \sqrt{\frac{n\log n}{2}}\right)\right) -  \sqrt{\frac{n\log n}{2}}} \geq 1-\eps/\tau - \frac{3}{n}.\]
After simplifying,
\[\PP{\sum_{i=1}^n P_i \geq  n \big( 1 - \alpha(1+\eps/\tau)(1-\gamma)\big) - \sqrt{2n \log n}} \geq 1-\eps/\tau - \frac{3}{n}.\]
Now we choose $\gamma = \frac{2.15}{\alpha}\sqrt{\frac{\log(n)}{n}}$ and $\tau=\eps n$
(we can assume that
$\gamma\leq 1$, since otherwise the theorem is trivial as it only claims that the coverage rate is no higher than 1). With this choice,
we calculate
\[ \PP{\sum_{i=1}^n P_i - \sqrt{\frac{n\log n}{2}} < 
(1-\alpha)(n+1)}\leq \frac{4}{n},\]
and so returning to our earlier calculations we have
\[\PP{Y_{n+1}\not\in\Ch_{n,\alpha}^{\textnormal{jackknife+},\eps}(X_{n+1})}  \geq 2\alpha(1-\gamma)\cdot \left(1 - \frac{5}{n}\right) 
\geq 2\alpha - 6\sqrt{\frac{\log(n)}{n}},\]
thus proving the theorem.

\end{document}